\documentclass[pra,aps,preprint,superscriptaddress,nofootinbib]{revtex4-2}
\usepackage{mathrsfs}
\usepackage{dcolumn}
\usepackage{bm}
\usepackage{hyperref}
\usepackage{epsfig,graphicx,color,appendix}
\usepackage{multirow}
\usepackage{tabularx}
\usepackage{subfigure}
\usepackage[countmax]{subfloat} 
\usepackage{threeparttable}
\usepackage{amssymb}
\usepackage{array}
\usepackage{amsmath}
\usepackage{makecell}
\usepackage{booktabs}
\usepackage{longtable}
\usepackage[dvipsnames]{xcolor}
\usepackage[normalem]{ulem}
\usepackage[markup=underlined]{changes}

\begin{document}

\title{Geometrical Tuning of Light-Matter Interaction in Atomic Trimer Antennas: A Symmetry-Resolved Modal Analysis}

\author{Linsa \surname{G. J.}}
\email{linsagj@cusat.ac.in}
\affiliation{International School of Photonics, Cochin University of Science and Technology, Kochi, Kerala, India, 682022.}

\author{Pushpender \surname{Singh}}
\email{pskkutlehria@gmail.com}
\affiliation{Department of Physics, Rabindranath Tagore Government College, Mandi, Himachal Pradesh, India, 175024.}

\author{Rohit \surname{Dhir}}
\email{dhir.rohit@gmail.com}
\affiliation{Department of Physics and Nanotechnology, SRM Institute of Science and Technology, Kattankulathur, Tamilnadu, India, 603203.}

\author{M. Ameen \surname{Poyli}}
\email[Corresponding author : ]{ameenpoyli@cusat.ac.in}
\affiliation{International School of Photonics, Cochin University of Science and Technology, Kochi, Kerala, India, 682022.}

\date{\today}

\begin{abstract}
Atomic trimers constitute the smallest geometry in which collective electric and magnetic responses emerge from coupled electric dipoles. We present a theoretical study of collective mode excitation in atomic trimers as the geometry is continuously tuned from linear to equilateral, using the coupled-dipole method with a multipole expansion formulated about the optimal scattering center. 
By combining eigenmode analysis and symmetry classification, we provide a complete symmetry-resolved map of the six in-plane and three out-of-plane modes, revealing how symmetry reduction across the $D_{\infty_h}$, $C_{2v}$, and $D_{3h}$ configurations governs the evolution of eigenmodes and their spectral features, lifting degeneracies, activating dark modes, and enabling full access to the modal spectrum. 
Based on this modal understanding, we demonstrate that forward-backward scattering can be switched solely by frequency detuning in a nearly linear trimer, without geometric reconfiguration. Furthermore, a linear trimer under $p$-polarized excitation supports a magnetic mode with a strongly enhanced magnetic field and a large Purcell factor, making it a promising platform for probing magnetic dipole transitions in atoms, with emission preferentially directed into the transverse plane. These results establish atomic trimers as a minimal platform where symmetry-controlled electric-magnetic mode engineering can be fully resolved and exploited for tailoring light-matter interaction at the atomic level. 
\end{abstract}

\maketitle

\section{INTRODUCTION}

Over the past decade, light scattering from single atoms, small atomic ensembles, and atoms coupled to subwavelength structures has been extensively studied within classical and quantum frameworks \cite{wang2010time, feng2013cooperative, lembessis2013two, lembessis2015radiation, bettles2015cooperative, bettles2016cooperative, asenjo2017atom}. Their interpretation as atomic antennas, together with the recognition of collective magnetic resonances, was established in 2020 \cite{alaee2020quantum, ballantine2020optical}. This reinterpretation initiated extensive research on few-emitter configurations such as dimers, trimers, and linear chains, both theoretically \cite{alaee2020quantum, alaee2020kerker, ballantine2020optical, parmee2022spontaneous} and experimentally \cite{rui2020subradiant, srakaew2023subwavelength}, with particular emphasis on their collective electric and magnetic responses. 
Although atoms are modeled as electric dipole emitters, their collective response supports induced currents that give rise to resonant electric and magnetic responses of comparable magnitude. Exploration of such modes has enabled the demonstration of the Kerker effect, superscattering, and scattering dark states \cite{alaee2020kerker}. These studies were carried out using relatively simple antenna configurations such as dimers, trimers, and linear chains, although more complex geometries have also been shown to support non-radiative anapole modes \cite{ballantine2020radiative}. 

The simplest form of an atomic antenna is a two-level atom, which acts as a naturally occurring electric dipole emitter \cite{scully1997quantum, zumofen2008perfect}. Single atoms, as well as few-atom configurations through their collective response, can serve as fundamental atomic antenna elements. Their point-like nature, spectrally narrow resonances arising from discrete transitions, and absence of ohmic losses are distinct advantages over plasmonic antennas. However, owing to the intrinsically narrow linewidths and limited tunability of atomic resonances, geometric control of the system parameters becomes essential for tailoring the optical response. The ability to detune the atomic resonance, together with the reconfigurable nature of atomic antennas, provides additional degrees of freedom \cite{sutherland2016collective, barredo2016atom, periwal2021programmable}. Together, these properties define the distinct advantages of atomic antennas. 

Collective responses extend to periodic arrays of atomic antennas, enabling atomic metasurfaces for wavefront control. Collective modes in such arrays have been explored for realizing electric and magnetic mirrors, Huygens' surfaces with beam steering, and generation of angular momentum beams \cite{bettles2016enhanced, shahmoon2017cooperative, ballantine2021cooperative, ballantine2022opticalSt}.
More recently, arrays of atomic emitters acting as metalenses have demonstrated the potential of atomic platforms for nanoscale light manipulation \cite{andreoli2025metalens}. 
Beyond these functionalities, atomic arrays have also been investigated as quantum metasurfaces for controlling polarization states of light \cite{wang2017design}, realizing linear optical elements \cite{bassler2023linear}, tunable directional emission \cite{fernandez2022directional}, and quantum applications including quantum information processing, generation of entangled photon states \cite{bekenstein2020quantum, ballantine2021quantum}, phase control of quantum photonic states \cite{lloyd2025single}, among others. These atomic configurations are experimentally realizable using optical tweezers \cite{barredo2014strong, barredo2016atom, barredo2018synthetic, Manetsch2025} and optical lattices \cite{rui2020subradiant, srakaew2023subwavelength}, which enable individual atoms to be trapped and positioned at well-defined and controllable locations. In the present work, the emitters are therefore treated as spatially localized atoms at prescribed positions, without assuming a material substrate or physical support; the analysis focuses on their collective electromagnetic response for a fixed geometry.

Among atomic antenna structures, triangular trimers are the smallest configuration that supports circulating current modes (e.g., an $\text{A}_2'$ mode with a net magnetic dipole moment), in addition to the electric and magnetic dipole responses found in dimers. Trimers offer a significantly richer set of modes that are tunable by geometric parameters, incident conditions, and detuning frequency. Previous studies of atomic trimers have largely focused on highly symmetric configurations (equilateral or linear), where modal degeneracies limit the accessible spectral features and obscure the full potential of geometric tuning. A systematic investigation of how collective modes evolve as the trimer geometry is continuously transformed from linear to equilateral, and how symmetry breaking enables otherwise dark modes, has remained absent. Furthermore, the optimal description of non-equilateral trimers requires a careful treatment of the multipole expansion about the true scattering center, which deviates from the geometric center when symmetry is reduced. 

Central to this study is the controlled reduction of symmetry: as the trimer geometry is tuned from linear ($D_{\infty h}$) through isosceles ($C_{2v}$) to equilateral ($D_{3h}$), modal degeneracies are lifted, previously dark modes become accessible, and the full in-plane and out-of-plane spectra emerge. In the present work, we address these gaps by performing a comprehensive theoretical analysis of geometrically tuned atomic trimer antennas. We employ the coupled-dipole method to describe collective light scattering and formulate the multipole expansion about the optimal scattering center obtained by minimizing higher-order moments. This refinement is essential for accurately capturing the response of isosceles and intermediate configurations. To interpret the scattering spectra, we further perform an eigenmode decomposition of the total response, which allows us to identify each spectral feature with a distinct collective mode. We begin with in-plane incidence, where the incident light is confined to the trimer plane. The trimer geometry is varied continuously from linear to equilateral, and the system is excited by $s$- and $p$-polarized plane waves, covering the full angular range required to access all symmetry-allowed modes. Through this approach, we map the evolution of the six in-plane and three out-of-plane modes, classify them by their symmetry point groups and irreducible representations (irreps), and track how the mode spectrum evolves with geometry. We then consider out-of-plane incidence, where the incident light is not confined to the trimer plane, and study the spectral evolution over the full range of incident angles.

Our analysis reveals how symmetry reduction controls mode excitation, including a crossing between two modes belonging to the same irrep. Based on the modal understanding, we demonstrate that forward-backward scattering can be switched solely by frequency detuning in a nearly linear trimer, without geometric reconfiguration. Additionally, we show that a linear trimer under $p$-polarized excitation supports a magnetic mode with a strongly enhanced magnetic field and a large Purcell factor, making it a promising platform for probing magnetic dipole transitions in atoms, with emission preferentially directed into the transverse plane, enabling efficient off-axis signal collection with reduced background. These results establish atomic trimers as a minimal platform where symmetry-controlled electric-magnetic mode engineering can be fully resolved and exploited for tailoring light-matter interaction at the atomic level. 

The paper is organized as follows. Section~\ref{sec:theory} describes the trimer geometry and theoretical framework, including the coupled-dipole method, the multipole expansion, and the eigenmode decomposition. Section~\ref{sec:GeomTune} presents the optical response under geometric tuning, evaluated about the shifted scattering center, for both $s$- and $p$-polarizations, first for in-plane incidence and then for out-of-plane incidence. Section~\ref{sec:FB} demonstrates detuning-controlled forward-backward switching. Section~\ref{sec:emission} analyzes the magnetic hotspot and Purcell enhancement. Section~\ref{sec:conclusion} concludes with an outlook. 

\section{SYSTEM AND THEORETICAL FRAMEWORK}
\label{sec:theory}
\subsection{Atomic trimer geometry}
\begin{figure} [h]
\centering
    \includegraphics[width = 1\textwidth,trim = 0pt 0pt 0pt 0pt, clip]{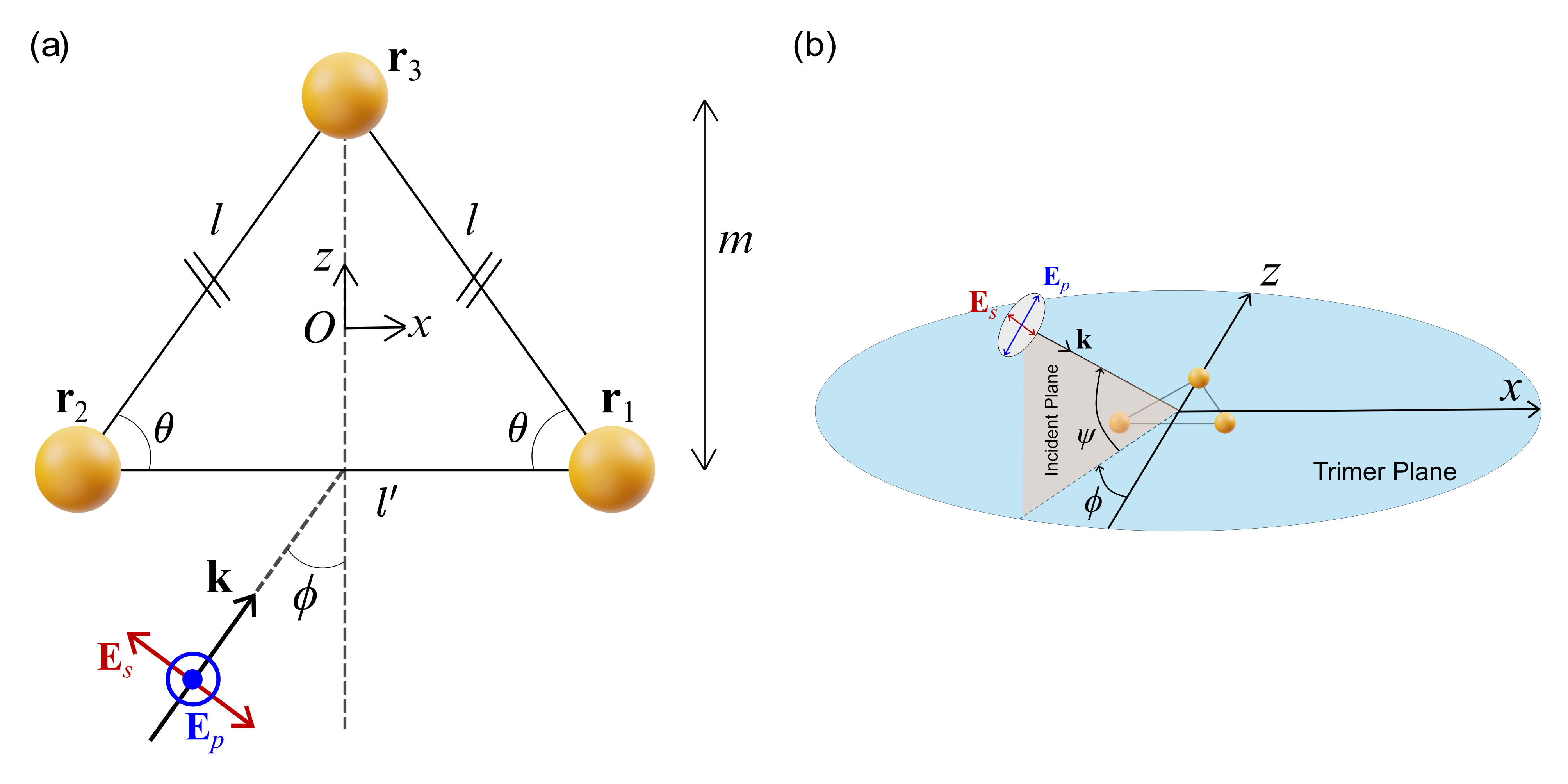}
    \caption{(a) Isosceles atomic trimer antenna consisting of three identical two-level atoms at positions $\mathbf{r}_{1}$, $\mathbf{r}_{2}$, and $\mathbf{r}_{3}$ in the $xz$ plane, with origin $O$. The leg length $l$, base length $l'$, and distance $m$ define the geometry, with $\theta$ being the base angle. The angle of incidence $\phi$, propagation vector $\mathbf{k}$, and polarization directions $\mathbf{E}_{s}$ and $\mathbf{E}_{p}$ for $s$- and $p$-polarized excitations are shown for in-plane incidence. (b) For out-of-plane incidence, the angle $\psi$ between $\mathbf{k}$ and the trimer plane is shown. The corresponding $\mathbf{E}_{s}$ and $\mathbf{E}_{p}$, together with the incident plane, are also indicated.}
\label{fig:Trimer_schematic}
\end{figure}

To investigate the optical response of geometrically tuned atomic antennas, we consider an atomic trimer consisting of three isolated identical two-level atoms, each modeled as an electric dipole emitter. The trimer has an isosceles geometry with atoms positioned at $\mathbf{r}_{1} = l'/2 \ \mathbf{e}_{x} - m/3 \ \mathbf{e}_{z}$, $\mathbf{r}_{2} = -l'/2 \ \mathbf{e}_{x} - m/3 \ \mathbf{e}_{z}$, and $\mathbf{r}_{3} = 2 m/3 \ \mathbf{e}_{z}$, as shown in Fig.~\ref{fig:Trimer_schematic}(a). The leg length $l$, base length $l'$, and the distance $m$ are marked in the figure, while $\mathbf{e}_{x}$ and $\mathbf{e}_{z}$ denote the unit vectors along the $x$ and $z$ axes, respectively. The isosceles trimer geometry varies continuously between the linear ($2l = l'$) and equilateral ($l = l'$) configurations as the base angle $\theta$ is tuned.\footnote{The prescribed geometry should be distinguished from a chemically bound molecular trimer, where internal coordinates can constitute dynamical vibrational degrees of freedom. Here, the interatomic separation and geometry are externally controlled.} 
The trimer lies in the $xz$ plane (trimer plane) with origin $O$, and its optical response is examined under both $s$- and $p$-polarized incident fields ($\mathbf{E}_s$ and $\mathbf{E}_p$, respectively). The trimer is illuminated by a linearly polarized plane wave $\mathbf{E}_\text{inc}(\mathbf{r}) = E_{0} \, e^{-i \mathbf{k} \cdot \mathbf{r}} \, \mathbf{e}$,  where $E_{0}$ is the field amplitude, $\mathbf{e}$ the polarization unit vector, and $\mathbf{k}$ the wave vector. When the incident wave vector lies entirely within the trimer plane, it is referred to as in-plane incidence. The corresponding in-plane angle $\phi$ is measured from the $z$-axis. To describe out-of-plane incidence, we introduce an out-of-plane angle $\psi$, defined as the angle between the incident wave vector and the trimer plane, as shown in Fig. \ref{fig:Trimer_schematic}(b). 
For $s$-polarization, the incident electric field is perpendicular to the incident plane, whereas for $p$-polarization, it lies within the incident plane. 
The incident plane and the directions of $\mathbf{E}_s$ and $\mathbf{E}_p$ are also shown. These parameters fully describe the linear, isosceles, and equilateral geometries for both in-plane and out-of-plane incidence with $s$- and $p$-polarized fields. The antenna dimensions are sub-wavelength, and each atom is modeled as an isotropic linear scatterer with polarizability  $\alpha(\omega_\text{inc}) = [-(\Gamma_{0}/2)\alpha_{0}]/[\delta + i (\Gamma_{0})/2]$, where $\alpha_{0} = 6\pi/k^{3}$, $\Gamma_{0}$ is the radiative decay-rate, and $\delta = \omega_\text{inc} - \omega_{a}$ is the frequency detuning between the incident frequency $\omega_\text{inc}$ and the atomic transition frequency $\omega_{a}$~\cite{lagendijk1996resonant, lambropoulos2007fundamentals}. Frequency detuning is normalized to $\Gamma_{0}$ and defined as $(\omega_\text{inc}-\omega_{a})/\Gamma_{0}$.\\

In an experimental realization, external electromagnetic confinement, such as an optical lattice or optical tweezers, would be required to localize the atoms and maintain their prescribed positions. These trapping fields are not explicitly included in the present model. Instead, we treat the atomic positions as fixed and calculate the collective electromagnetic response to the incident probe field within the linear coupled-dipole approximation. The incident-field amplitude ($E_0$) is normalized to unity in our calculation and does not correspond to a specific absolute experimental probe intensity.

For an experimental reference, we consider the optical-lattice implementation involving $^{87}\mathrm{Rb}$ atoms reported in Ref.~\cite{rui2020subradiant}. Using the reported lattice depth and the photon fluence for the reflection measurement (approximately 27 photons per site over 5 ms) \cite{rui2020subradiant}, we estimate a probe intensity of $\sim 4.50\times10^{-3}~\mathrm{W\,m^{-2}}$. The corresponding trapping intensity is $\sim 2.03\times10^{8}~\mathrm{W\,m^{-2}}$ per lattice axis, yielding an indicative trap-to-probe intensity ratio of $\sim 4.51\times10^{10}$. The very small probe intensity relative to the trapping intensity is consistent with operation in the weak-excitation regime. In a recent study, collective enhancement and suppression of light scattered into an optical cavity were also observed with tweezer-trapped $^{87}\mathrm{Rb}$ arrays under low-saturation conditions \cite{yan2023superradiant}.

The above intensity comparison provides an experimental reference for the ratio between trapping and probe intensities and supports treating the probe as a weak perturbation to the confined atomic configuration. These parameters are adopted solely as intensity benchmarks. The intensity ratio does not by itself establish that trapping-induced modifications of the atomic transition frequencies are negligible; such modifications, including AC Stark shifts, are not incorporated in the present model, which assumes identical, isotropic two-level atoms. In a physical realization, such trap-induced effects and unwanted excitations can be reduced by suitably employing far-detuned trapping light, with the choice of trapping wavelength depending on the atomic species and the transitions involved \cite{rui2020subradiant, yan2023superradiant, miller1993far, jun2008quantum, GRIMM200095}.

\subsection{Theoretical framework}
\subsubsection{\textbf{Multipole expansion of the trimer response}}
The optical response of the atomic trimer antenna is governed by collective dipole-dipole interactions between the constituent atoms. Within the coupled-dipole framework, a Green's tensor is constructed for the trimer, wherein each atom responds to the superposition of the incident field and the fields scattered by the neighboring atoms. The induced dipole moment $\mathbf{p}(\mathbf{r}_{i})$ of each atom ($i=1,2,3$) at position $\mathbf{r}_{i}$ is then given by~\cite{foldy1945multiple, mulholland1994light} 
\begin{equation} 
    \label{eq:CDM} 
    \mathbf{p}(\mathbf{r}_{i}) = \epsilon_{0} \alpha \left[\mathbf{E}_\text{inc}(\mathbf{r}_{i}) + \sum_{j=1, j \neq i}^{3} \mathbf{G}(\mathbf{r}_{i}, \mathbf{r}_{j}) \mathbf{p}(\mathbf{r}_{j}) \right], 
\end{equation} 

where $\mathbf{G}(\mathbf{r}_{i}, \mathbf{r}_{j})$ is the Green's tensor mediating the dipole-dipole interaction between atoms at $\mathbf{r}_{i}$ and $\mathbf{r}_{j}$; we adopt the standard free-space dyadic formulation provided in Appendix \ref{App.A}.

The individual dipole moment components $p_{i}^{x}$, $p_{i}^{y}$, and $p_{i}^{z}$, obtained by solving Eq.~\eqref{eq:CDM} for each atom, serve as the input to the multipole decomposition~\cite{alaee2018electromagnetic}, from which we derive the effective induced electric and magnetic dipole moments of the trimer. 
The governing equations for the isosceles trimer under both $s$- and $p$-polarized incidence are given as follows:

\begin{equation} \label{eq:pxeff}
\begin{split}
    p_{\mathrm{eff}}^{x} = \sum_{i=1}^{3} \Bigg[
    p_i^{x} \, j_0\!\left(k |\mathbf{r}_i|\right) + \frac{3k^{2}}{2} \left( p_i^{x} (r_i^{x})^{2} + p_i^{z} r_i^{x} r_i^{z} \right) \, \frac{j_2\!\left(k |\mathbf{r}_i|\right)}{\left(k |\mathbf{r}_i|\right)^{2}} - \frac{k^{2}}{2} \, p_i^{x} |\mathbf{r}_i|^{2} \, \frac{j_2\!\left(k |\mathbf{r}_i|\right)}{\left(k|\mathbf{r}_i|\right)^{2}} \Bigg],
    \end{split}
\end{equation}
\begin{equation}\label{eq:pyeff}
    \begin{split}
        p^{y}_\text{eff} = \sum\limits_{i=1}^{3} \Bigg[ p^{y}_{i} \, j_{0}\!\left(k|\mathbf{r}_{i}|\right) - \frac{k^{2}}{2} p^{y}_{i} |\mathbf{r}_{i}|^{2} \, \frac{j_{2}\!\left(k |\mathbf{r}_{i}|\right)}{(k|\mathbf{r}_{i}|)^{2}} \Bigg],
    \end{split}
\end{equation}
\begin{equation}\label{eq:pzeff}
    \begin{split}
    p^{z}_\text{eff} = \sum\limits_{i=1}^{3} \Bigg[ p^{z}_{i} \, j_{0} \!\left(k|\mathbf{r}_{i}|\right) + \frac{3 k^{2}}{2} (p^{z}_{i} (r^{z}_{i})^{2} + p^{x}_{i} r^{x}_{i} r^{z}_{i}) \, \frac{j_{2}\!\left(k |\mathbf{r}_{i}|\right)}{(k|\mathbf{r}_{i}|)^{2}} - \frac{k^{2}}{2} p^{z}_{i} |\mathbf{r}_{i}|^{2} \, \frac{j_{2}\!\left(k |\mathbf{r}_{i}|\right)}{(k|\mathbf{r}_{i}|)^{2}} \Bigg],
    \end{split}
\end{equation}
\begin{equation}\label{eq:mxeff}
    m^{x}_\text{eff} = \frac{3 i \omega}{2} \sum\limits_{i=1}^{3} \Bigg[ (p^{y}_{i} r^{z}_{i}) \, \frac{j_{1}\!\left(k|\mathbf{r}_{i}|\right)}{k |\mathbf{r}_{i}|} \Bigg], 
\end{equation}
\begin{equation}\label{eq:myeff}
    m^{y}_\text{eff} = -\frac{3 i \omega}{2} \sum\limits_{i=1}^{3} \Bigg[ (p^{x}_{i} r^{z}_{i} - p^{z}_{i} r^{x}_{i}) \, \frac{j_{1}\!\left(k|\mathbf{r}_{i}|\right)}{k |\mathbf{r}_{i}|} \Bigg], 
\end{equation}
\begin{equation}\label{eq:mzeff}
    m^{z}_\text{eff} = -\frac{3 i \omega}{2} \sum\limits_{i=1}^{3} \Bigg[ (p^{y}_{i} r^{x}_{i}) \, \frac{j_{1}\!\left(k|\mathbf{r}_{i}|\right)}{k |\mathbf{r}_{i}|} \Bigg], 
\end{equation}
    
where $j_{0}$, $j_{1}$, and $j_{2}$ are the spherical Bessel functions of respective orders, and $p^{x}_\text{eff}$, $p^{y}_\text{eff}$, $p^{z}_\text{eff}$, $m^{x}_\text{eff}$, $m^{y}_\text{eff}$, $m^{z}_\text{eff}$ are the components of the effective electric and magnetic dipole moments of the antenna. The individual components $p_{i}^{x}$, $p_{i}^{y}$, and $p_{i}^{z}$ are derived from the coupled-dipole model for an arbitrary angle of incidence (see Appendix~\ref{App.A}). For $s$-polarization, only $p^{x}_\text{eff}$, $p^{z}_\text{eff}$, and $m^{y}_\text{eff}$ are non-vanishing, whereas $p$-polarization can excite all six components, with $p^{y}_\text{eff}$, $m^{x}_\text{eff}$, and $m^{z}_\text{eff}$ additionally present depending on the incidence. Once the effective multipole moments are evaluated, the total scattering cross section is obtained as 
\begin{equation} \label{eq:MPD}
    C_{\text{sca}} = \frac{k^{4}}{6 \pi \epsilon_{0}^{2} E_{0}^{2}}  \sum\limits_{\alpha} \Bigg[ \left( p^{\alpha}_{\text{eff}} \right)^{2} + \left(m^{\alpha}_\text{eff}/c \right)^{2} \Bigg],
\end{equation} where $\alpha \in \{x,y,z\}$. 
    
\subsubsection{\textbf{Eigenmode decomposition}} \label{sec:EMD}

The electromagnetic response of trimer antennas can be interpreted as the superposition of excited resonant eigenmodes.
Eigenmode studies on linear arrangements of atoms have been carried out  to identify and visualize the collective modes \cite{bettles2016cooperative}. 
We employ eigenmode decomposition of the trimer antennas to understand the physical origin and  characteristics of their fundamental collective modes.
This allows us to quantify their individual contributions to the total response of the antenna for a given geometry. 
The total scattering cross section ($\sigma_{sc}$) of an ensemble of electric dipoles, from eigenmode decomposition, is given by~\cite{jackson1998classical, bohren1983absorption, novotny2012principles} 
\begin{equation} \label{eq:EMD}
    \sigma_{sc} = \frac{\sigma_{0}}{\alpha_{0} |E_\text{inc}|^{2}} \left [\sum_{n} |b_{n}|^{2} \, \text{Im} \left (\frac{1}{\mu_{n}}\right ) + \sum_{n,n'}^{n \neq n'} \text{Im}\left(\frac{b_{n}^{*} b_{n'}}{\mu_{n'}} \, \mathbf{m}_{n}^{*} \cdot \mathbf{m}_{n'}\right) \right].
\end{equation}
Here, $ \sigma_{0} = 6 \pi / k^{2}$ is the resonant atomic scattering cross section. 
$\mu_{n}$ and $\mathbf{m}_{n}$ are the eigenvalues and eigenvectors of the coupling matrix $M$, respectively (Appendix \ref{App.B}). 
These characterize the collective excited modes for an $N$-atom system, each labeled by an eigenmode index $n \in \{ 1,2,...,3N \}$. For an atomic trimer, $M$ is a $9 \times 9$ complex symmetric matrix with nine eigenvectors and eigenvalues. 
The first term in Eq.~(\ref{eq:EMD}) with the sum over $n$, accounts for the individual modal contributions to scattering, while the second sum represents the interference contributions between different modes. 

\section{\MakeUppercase{Geometry-Controlled Symmetry Breaking and Mode Access in Atomic Trimers}}
\label{sec:GeomTune}
	
\begin{figure} [h]
    \centering
    \includegraphics[width = \textwidth]{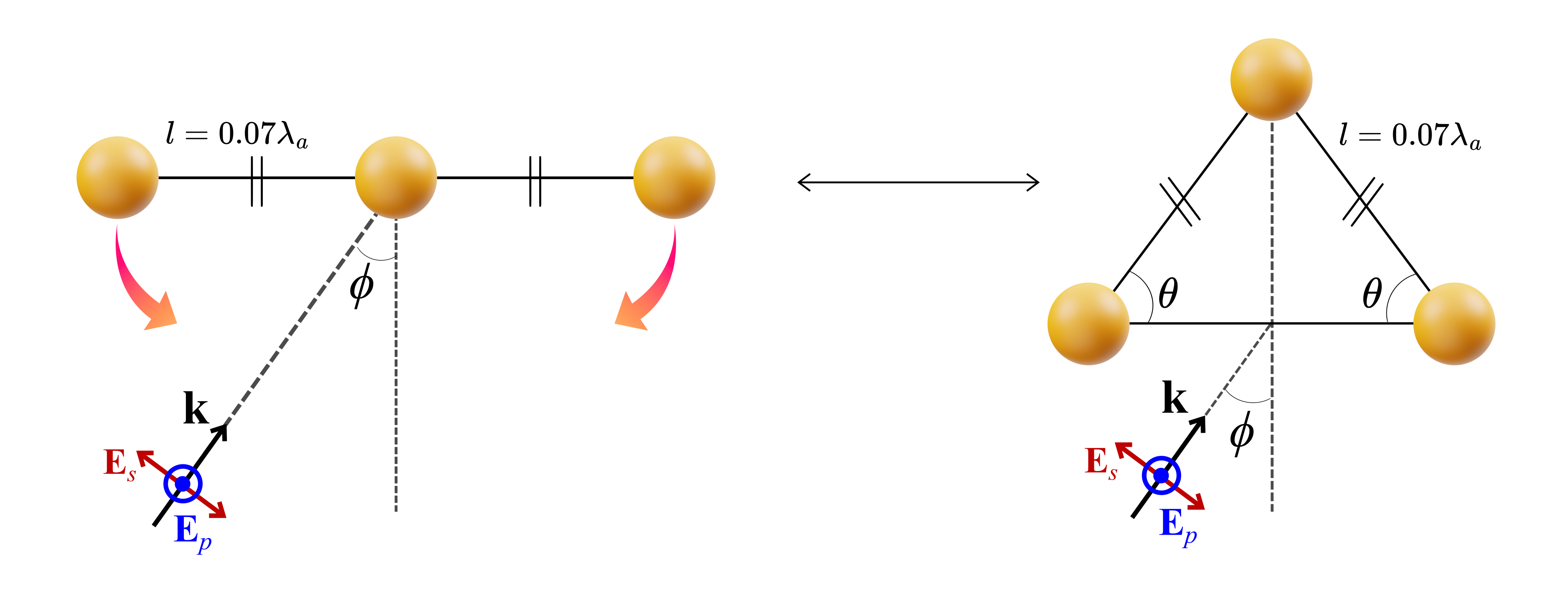}
    \caption{Schematic representation of tuning the geometry of the trimer antenna under in-plane excitation for $s$- and $p$-polarizations ($\mathbf{E}_s$ and $\mathbf{E}_p$, respectively).
    A linear trimer with fixed interatomic distance is reshaped to form an isosceles triangular configuration eventually ending up in an equilateral trimer. 
    The interatomic distance in the linear trimer and the leg length of the isosceles trimer are both fixed at $l=0.07\lambda_a$. The base angle $\theta$ and the in-plane angle of incidence $\phi$ are marked.}
    \label{fig:Trimer_tuning}
\end{figure}

Atomic antennas of triangular and linear geometry have been studied for light-matter  interaction at the atomic scale. Systems such as equilateral trimers and linear chains are well understood with aims of obtaining optical magnetism, directional scattering, and scattering dark states \cite{alaee2020kerker,parmee2022spontaneous}. Such studies mainly focused on highly symmetric systems which excite only certain fundamental modes. However, controlling light-matter interaction via symmetry breaking, enabling access to the full set of modes, remains largely unexplored. In principle, all nine modes can be excited under plane wave illumination with $s$- and $p$-polarized incidence. We therefore focus on configurations intermediate between linear and equilateral trimers that possess additional spectral features arising from new modes excited due to symmetry breaking, leading to a spectrally rich scattering response. The presence of new spectral features demands a full eigenmode analysis to understand the nature of the newly excited modes, resolve their individual contributions, and thus establish their role in the overall antenna response. Gaining control over mode excitation via geometrical tuning is essential for understanding and tailoring the antenna response, including directional scattering, selective mode excitation, and switchable optical functionalities. Exploiting these responses can lead to applications including atom-level sensing, single atom spectroscopy, and the design of tunable photonic devices.\\
\indent
We consider an isosceles trimer with leg lengths fixed at $l = 0.07 \lambda_a$, $\lambda_a$ being the atomic transition wavelength. 
The optical response of the system is studied for various base angles $\theta$ as its geometry transforms from a linear ($\theta =0^\circ$) to an equilateral ($\theta =60^\circ$) configuration, as schematically shown in Fig. \ref{fig:Trimer_tuning}, for in-plane incidence. 
The fixed interatomic distance of the linear trimer becomes the leg length of the isosceles trimer as the geometry transforms into a triangular shape.

\indent
For an equilateral trimer, the scattering center coincides with the geometric center of the antenna~\cite{ustimenko2025optimal}. 
However, when the antenna is tuned away from the equilateral configuration, the scattering center deviates from the geometric center. 
When multipole decomposition is performed, this effect must be incorporated into the derivation of the effective dipole moments to accurately describe the response of antennas with an isosceles geometry. 
One approach to find the optimal scattering center is to minimize the effective higher-order moments \cite{kildishev2025art,ospanova2023modified}, thereby obtaining the scattering center where the dipole contribution is maximal and higher-order moments vanish. 
We therefore minimize the effective electric and magnetic quadrupolar moments of the trimer and derive the positions of the shifted electric and magnetic scattering centers. 
The modified multipole moments thus obtained for $s$- and $p$-polarizations are presented in the following sections. 

\subsection{In-plane collective modes} \label{subsec:InplaneModes}
We first excite the triangular trimer with an $s$-polarized plane wave in order to access the six in-plane modes. The incident wave vector is confined to the trimer plane. Since $s$-polarized incidence induces electric dipole moments within the trimer plane, the resulting modes are referred to as in-plane modes. For $s$-polarization, the modified multipole moments $p'^{x}_\text{eff}$, $p'^{z}_\text{eff}$, and $m'^{y}_\text{eff}$, obtained by incorporating the effects of the shifted scattering centers into Eqs. (\ref{eq:pxeff}), (\ref{eq:pzeff}), and (\ref{eq:myeff}) are given below:
\begin{equation} \label{eq:pxeffp}
    \begin{split}
        p'^{x}_\text{eff} = \sum\limits_{i=1}^{3} \Bigg[ p^{x}_{i} \, j_{0} \!\left(k|\mathbf{r} - \mathbf{d}_{e}|_{i}\right) + \frac{3 k^{2}}{2} (p^{x}_{i} (r^{x} - d^{x}_{e})_i^{2} + p^{z}_{i} (r^{x} - d^{x}_{e})_{i} (r^{z} - d^{z}_{e})_{i}) \\ \frac{j_{2}\!\left(k |\mathbf{r}-\mathbf{d}_{e}|_{i} \right)}{(k|\mathbf{r} - \mathbf{d}_{e}|_{i})^{2}} - \frac{k^{2}}{2} p^{x}_{i} |\mathbf{r} - \mathbf{d}_{e}|_{i}^{2} \, \frac{j_{2}\!\left(k |\mathbf{r} - \mathbf{d}_{e}|_{i}\right)}{(k|\mathbf{r}-\mathbf{d}_{e}|_{i})^{2}} \Bigg],
    \end{split}
\end{equation}
\begin{equation} \label{eq:pzeffp}
    \begin{split}
        p'^{z}_\text{eff} = \sum\limits_{i=1}^{3} \Bigg[ p^{z}_{i} \, j_{0} \!\left(k|\mathbf{r}-\mathbf{d}_{e}|_{i}\right) + \frac{3 k^{2}}{2} (p^{z}_{i} (r^{z} - d^{z}_{e})_{i}^{2} + p^{x}_{i} (r^{x} - d^{x}_{e})_{i} (r^{z}-d_{e}^{z})_{i}) \\  \frac{j_{2}\!\left(k |\mathbf{r}-\mathbf{d}_{e}|_{i}\right)}{(k|\mathbf{r} - \mathbf{d}_{e}|_{i})^{2}} 
        - \frac{k^{2}}{2} p^{z}_{i} |\mathbf{r} - \mathbf{d}_{e}|_{i}^{2} \, \frac{j_{2}\!\left(k |\mathbf{r}-\mathbf{d}_{e}|_{i}\right)}{(k|\mathbf{r}-\mathbf{d}_{e}|_{i})^{2}} \Bigg],
    \end{split}
\end{equation}
\begin{equation} \label{eq:myeffp}
    \begin{split}
        m'^{y}_\text{eff} = -\frac{3 i \omega}{2} \sum\limits_{i=1}^{3} \Bigg[ (p^{x}_{i} (r^{z}-d_{m}^{z})_{i} - p^{z}_{i} (r^{x}-d_{m}^{x})_{i}) \, \frac{j_{1}\!\left(k|\mathbf{r}-\mathbf{d}_{m}|_{i}\right)}{k |\mathbf{r}-\mathbf{d}_{m}|_{i}} \Bigg], 
    \end{split}
\end{equation}	
where $d_{e}^{x}, d_{e}^{z}, d_{m}^{x}$, and $d_{m}^{z}$ are the shifts of the electric ($e$) and magnetic ($m$) moments from the geometric center along the respective directions.
These modified moments $p'^{x}_\text{eff}$, $p'^{z}_\text{eff}$, and $m'^{y}_\text{eff}$ are then substituted into Eq.~(\ref{eq:MPD}) for a more accurate calculation of the scattering cross section. 

\subsubsection{\textbf{Normal incidence}}	\label{subsubsec:NormInc}

\begin{figure}
    \centering
    \includegraphics[width = 1\textwidth,trim = 1pt 0pt 0pt 0pt, clip]{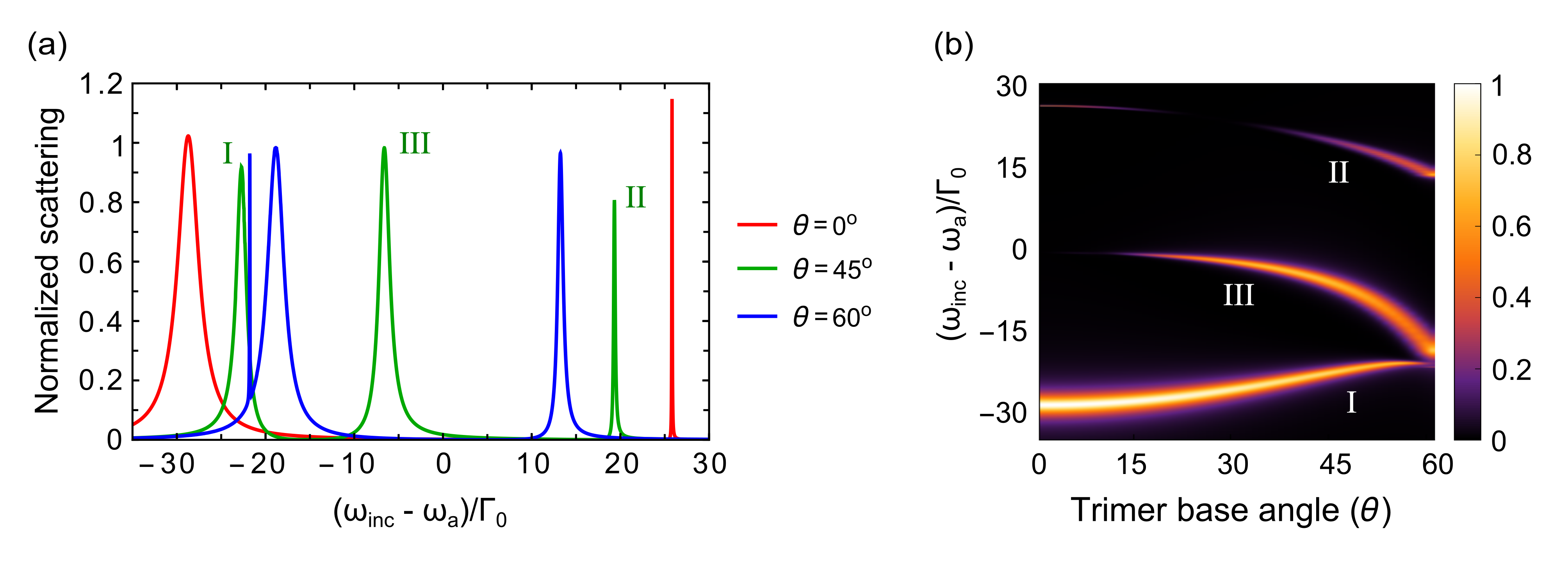}
    \caption{Geometric tuning of the optical response of a trimer under in-plane $s$-polarized normal excitation. (a) Normalized scattering cross section $C_\text{sca}({3 \lambda^{2}}/{2 \pi})$ for linear (red), isosceles ($\theta = 45^\circ$, green) and equilateral (blue) trimers, plotted as a function of detuning frequency.
    The linear trimer shows two peaks, whereas an additional peak is present in both triangular geometries. 
    (b) Map of the scattering cross section showing the evolution of the spectral peaks (labeled I, II and III) as the trimer is transformed from linear to equilateral configuration as a function of $\theta$. The leg length is fixed at $l = 0.07 \lambda_{a}$ (see Fig. \ref{fig:Trimer_tuning}).}
    \label{fig:trimer_scat}
\end{figure}

We begin with normal incidence ($\phi=0^\circ$), where symmetry yields a simpler spectrum due to modal degeneracies. Total scattering by the antenna is calculated for different configurations ranging from linear to equilateral geometry. Figure~\ref{fig:trimer_scat}(a) compares the total scattering, normalized to $3\lambda^{2}/2\pi$, as a function of frequency detuning for three specific configurations: linear ($\theta = 0^\circ$, red), isosceles ($\theta = 45^\circ$, green), and equilateral ($\theta = 60^\circ$, blue). The interatomic distances in the linear case and the leg lengths in the isosceles case are both fixed at $l = 0.07 \lambda_{a}$, as shown in Fig.~\ref{fig:Trimer_tuning}. For the linear configuration, only two peaks  appear in the spectra, as the number of modes in this configuration for normal incidence is restricted to two \cite{feng2013cooperative}. As the linear configuration is reshaped into an isosceles one, the existing peaks shifts spectrally toward each other, and an additional peak appears in between them. This is shown for a base angle of $\theta = 45^\circ$ (green), where the existing peaks are labeled  I and II, and the new peak is labeled III. As the antenna is further deformed by increasing the base angle $\theta$ beyond $45^\circ$, peaks I and II move closer to each other. Additionally, the new peak III continues to redshift and eventually nearly merges with peak I as the equilateral configuration is attained.

\begin{figure}[h!]
    \centering
    \includegraphics[width = 0.55\textwidth]{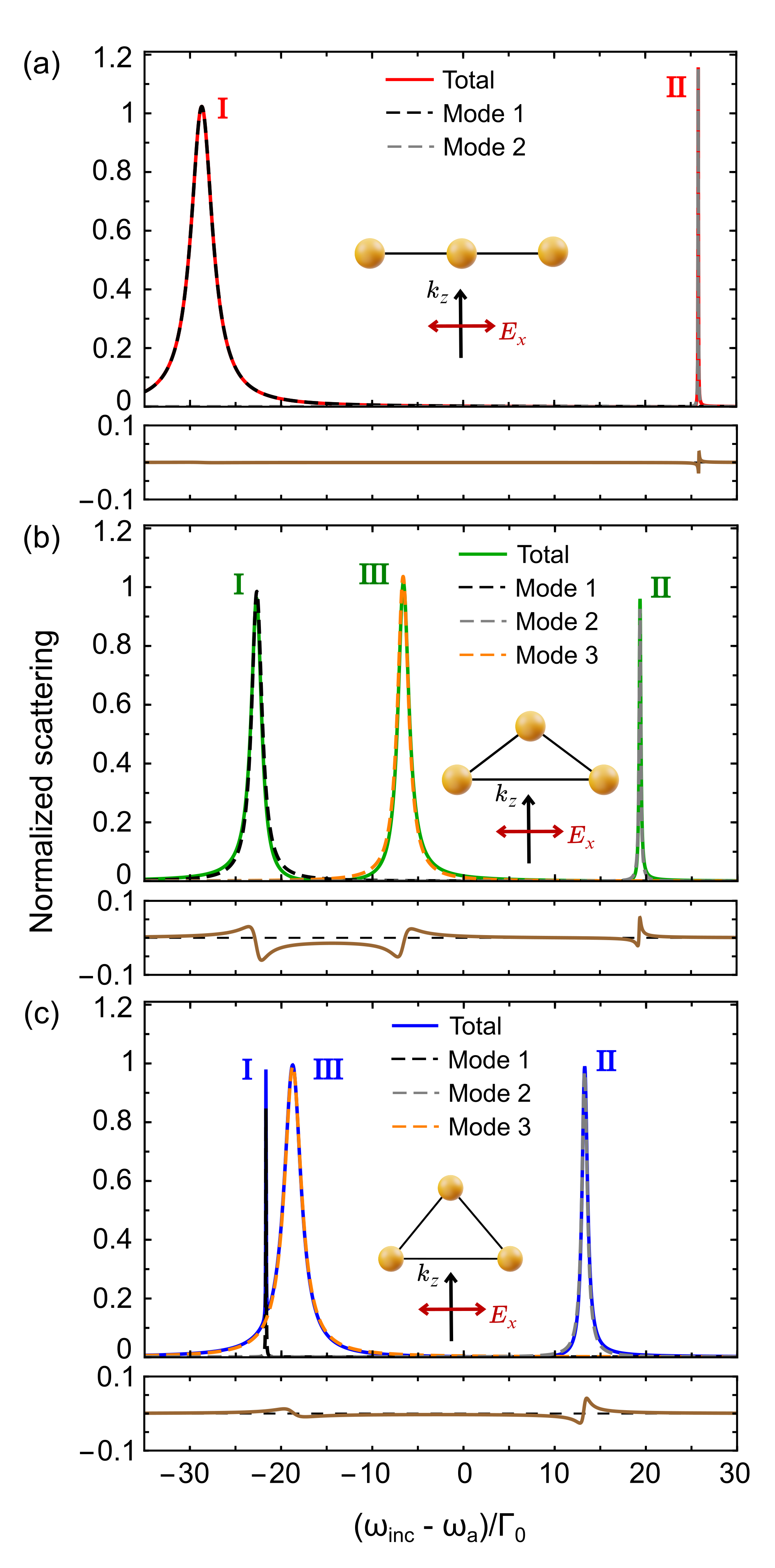}
    \caption{Eigenmode decomposition of atomic trimer with $l=0.07\lambda_{a}$ for in-plane $s$-polarized normal incidence. Modal contributions to the total scattering cross section are plotted as a function of detuning frequency for geometries: (a) linear, (b) isosceles ($\theta = 45^\circ$), and (c) equilateral. Total scattering is shown using solid lines, while modal contributions to peaks I, II, and III are shown with dashed lines. The interference contributions are shown separately in brown.}
    \label{fig:eigen_scat}
\end{figure}

To better understand the evolution of the peaks, we calculate the scattering cross section as a function of detuning while gradually varying the trimer base angle from $0^\circ$ to $60^\circ$. The result is shown in Fig. \ref{fig:trimer_scat}(b), where the shift of the scattering peaks with detuning and the merging of peaks I and III are clearly visible. The intensity of peak II is reduced in the geometric range $20^\circ \leq \theta \leq 30^\circ$. The strength and evolution of the observed peaks, governed by the change in interatomic coupling with geometric tuning, can be understood from the interaction between the excited fundamental modes. To clearly identify these modes, we perform a full eigenmode decomposition of all antenna modes, and the results are shown in Fig.~\ref{fig:eigen_scat}. 

The eigenmode analysis reveals that a triangular trimer supports nine modes in total, six of which are in-plane modes. Since we restrict to in-plane $s$-polarized illumination, only the six in-plane modes can be excited. Their strength depends on the antenna configuration and the incident angle. In the present case of normal incidence, only three fundamental modes are excited. Figure \ref{fig:eigen_scat} shows how the combined effect of these modes gives rise to the three observed spectral peaks. The modal contributions to the total scattering for the three configurations considered (linear, isosceles ($\theta = 45^\circ$), and equilateral) are shown. The solid curves represent the total scattering, while the contributions from each mode are shown with dashed curves. Figure \ref{fig:eigen_scat}(a) shows that the scattering from the linear trimer results from only two spectrally well separated modes 1 and 2, shown with black and gray dashed curves, respectively. The new peak appearing in the isosceles trimer results from the excitation of a new mode (mode 3), as shown in Fig. \ref{fig:eigen_scat}(b) with an orange dashed curve. Upon reaching the equilateral geometry, modes 1 and 3 become much closer to each other, as shown in Fig. \ref{fig:eigen_scat}(c). The interference contribution (second term in Eq. (\ref{eq:EMD}), shown in brown) is negligible.

Although the preceding analysis identifies the modes, understanding their nature and how they change with antenna configuration is essential for explaining the strength and evolution of the spectral peaks. 
To this end, using multipole decomposition and eigenmode analysis, we calculate the strength and orientation of each atomic dipole and examine how they evolve as the antenna configuration is tuned. 
The results are summarized in Table \ref{tab:0.07_modes}, along with their corresponding symmetry point groups and irreps, classified following the standard group-theoretical methods \cite{cotton1990chemical}, and symmetry-based analyses of trimer systems \cite{chuntonov2011trimeric, qiu2018symmetry}. 
This analysis provides insight into the evolution and transformation pathways of the excited modes as the trimer geometry is tuned. 
The arrows representing the dipole directions illustrate the nature and symmetry of the modes, while their lengths qualitatively indicate the relative strengths.
\begin{table}
    \centering
    \caption{Mode classification for trimer geometries ranging from linear to equilateral under in-plane normal incidence: all modes contributing to the three spectral peaks in Fig. \ref{fig:eigen_scat} are categorized according to their respective symmetry point groups and irreps. The linear, isosceles and equilateral trimer configurations belong to the $D_{\infty h}$, $C_{2v}$ and $D_{3h}$ symmetry point groups, respectively. The dipoles form a collinear bonding and antibonding arrangements at peak I and II, respectively, in the linear configuration and gradually evolve into two distinct circular current modes as the equilateral geometry is attained. Mode 3 contributing to peak III has a relatively weak central dipole. Bonding mode 1 evolves as $\Sigma_u^+ \rightarrow \text{B}_1 \rightarrow \text{A}_2'$; antibonding mode 2 evolves as $\Sigma_u^+ \rightarrow \text{B}_1 \rightarrow \text{E}'$; and mode 3 evolves as $\text{B}_1 \rightarrow \text{E}'$, undergoing a phase-reversal of outer dipoles near $\theta \approx 29^\circ$.}
    \includegraphics[width = 0.7\textwidth]{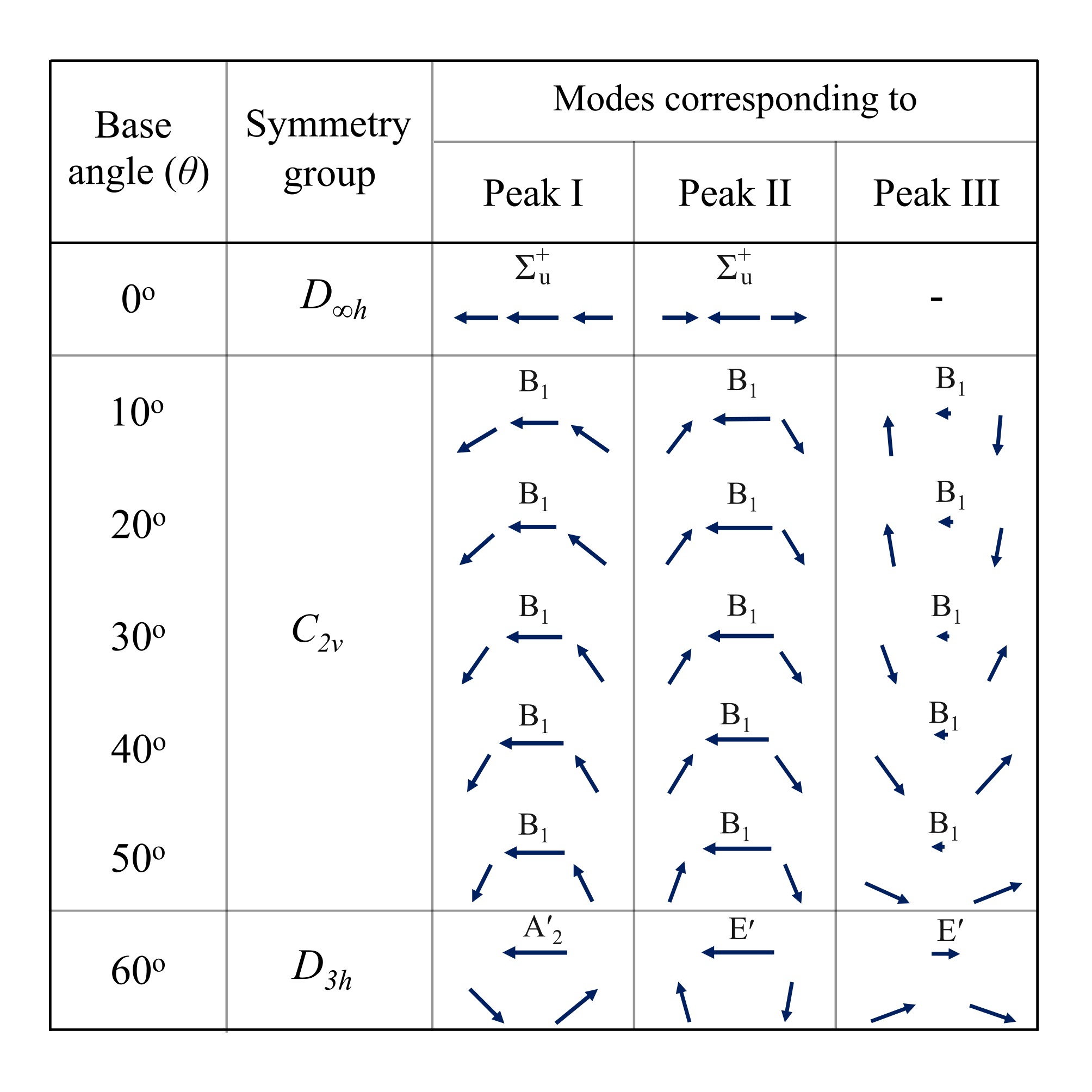}
    \label{tab:0.07_modes}
\end{table}
    
\begin{figure}[h!]
    \centering
    \includegraphics[width = 0.8\textwidth,trim = 0pt 0pt 0pt 0pt, clip]{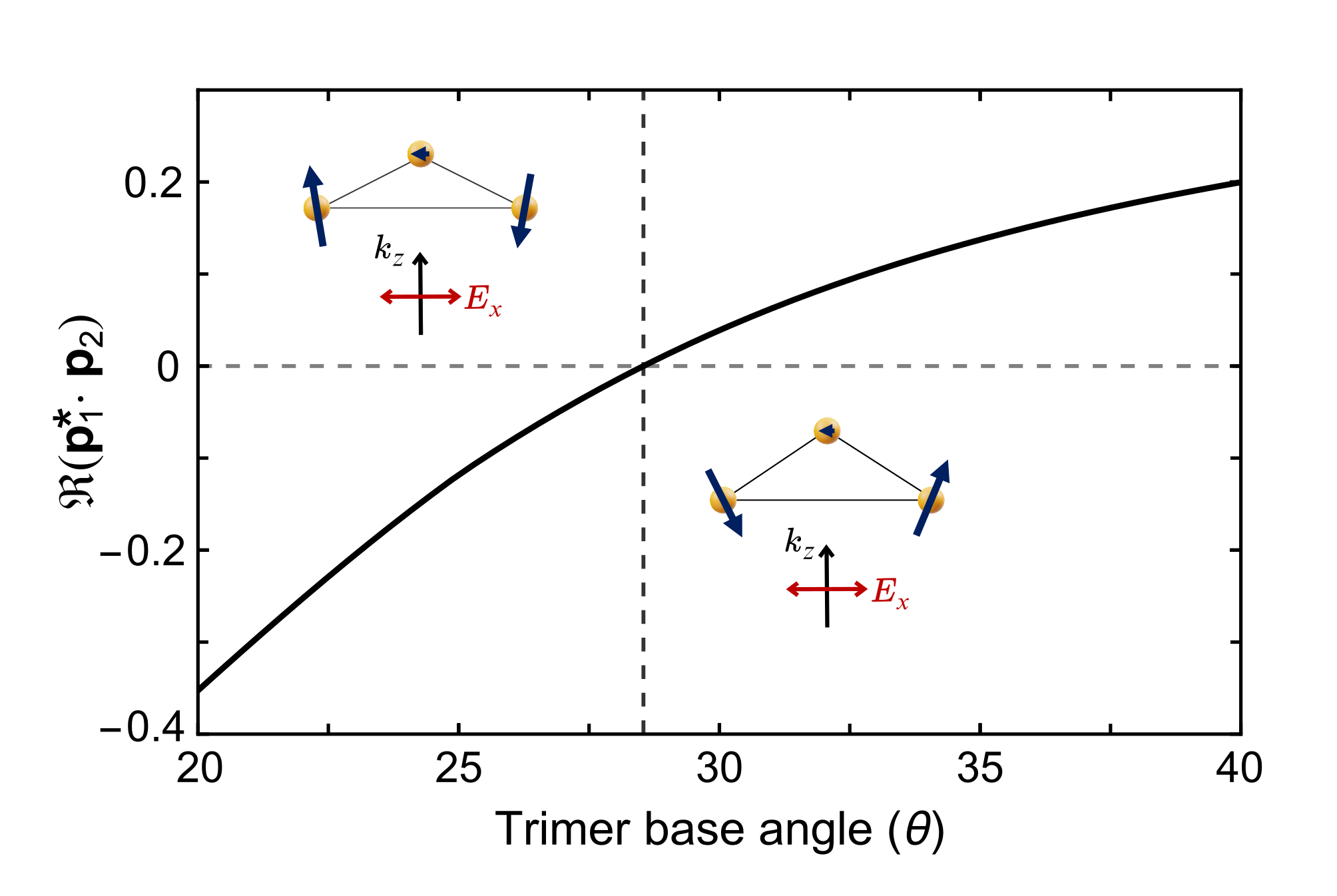}
    \caption{Phase reversal of the outer dipoles of mode 3 in an isosceles trimer excited under in-plane $s$-polarized normal incidence. The phase reversal is evident from $\Re(\mathbf{p}_{1}^{*} \cdot \mathbf{p}_{2})$ plotted as a function of the base angle $\theta$, with the dashed lines marking the zero crossing at $\theta \approx 29^\circ$. The corresponding dipole orientations on either sides of the zero crossing illustrate the phase reversal.}
    \label{fig:Phase_rev}
\end{figure}

A linear atomic trimer belongs to the $D_{\infty h}$ symmetry point group. 
Under normal incidence, the system supports two collinear modes - one bonding and one antibonding in nature - shown as mode 1 and mode 2 in Fig. \ref{fig:eigen_scat}(a), respectively, which transform according to the $\Sigma_u^+$ irrep. When the trimer is deformed into an isosceles geometry, lowering the symmetry to the $C_{2v}$ point group, the collinear modes begin to round up, and both modes transform according to the $\text{B}_1$ irrep. When the trimer attains equilateral geometry, belonging to the highly symmetric $D_{3h}$ point group, the bonding $\text{B}_1$ mode 1 associated with spectral peak I evolves into the well-defined $\text{A}'_2$ irrep, while the antibonding $\text{B}_1$ mode 2 associated with spectral peak II evolves into a doubly degenerate $\text{E}'$ irrep. The $\text{A}'_2$ mode exhibits a magnetic character arising from its bonding circular current configuration, while the $\text{E}'$ mode remains nonmagnetic due to its antibonding character. The third mode, corresponding to the additional peak III in the isosceles configuration (mode 3 in Fig. \ref{fig:eigen_scat}(b)), also belongs to the $\text{B}_1$ irrep; when the trimer attains equilateral geometry, it evolves into a doubly degenerate $\text{E}'$ irrep. The strengths of its individual dipoles differ significantly, with a relatively weak middle dipole, as shown in the last column of Table \ref{tab:0.07_modes}, in contrast to modes 1 and 2, where the strengths are comparable. The system exhibits an antibonding character for smaller base angles, and a phase reversal of the outer dipoles occurs as $\theta$ approaches $30^\circ$, switching the dipole configuration to a bonding nature, preserving its $\text{B}_{1}$ symmetry. This is quantified by the negative real part of the scalar product between the outer dipole moments $\Re(\mathbf{p}_{1}^{*} \cdot \mathbf{p}_{2}) < 0$, at smaller base angles. Calculations with varying $\theta$ show that this correlation changes sign at $\theta \approx 29^\circ$, marking a transition to a bonding-like phase relation. This behaviour is illustrated in Fig.~\ref{fig:Phase_rev}, where $\Re(\mathbf{p}_{1}^{*} \cdot \mathbf{p}_{2})$ is plotted as a function of $\theta$ for mode 3. The dashed lines indicate the zero crossing at $\theta \approx 29^\circ$, while the corresponding dipole configurations on either side of the crossing explicitly visualize the reversal of their relative phase relation. Peak III is not solely associated with mode 3; it also contains contributions from other modes, the relative weights of which determine the relative phases, and thus the orientations of the dipoles. To summarize, the geometric tuning of the atomic trimer from linear to equilateral configuration reveals that bonding mode 1 evolves as $\Sigma_u^+ \rightarrow \text{B}_1 \rightarrow \text{A}_2^{'}$, antibonding mode 2 evolves as $\Sigma_u^+ \rightarrow \text{B}_1 \rightarrow \text{E}'$, and mode 3 evolves as $\text{B}_1 \rightarrow \text{E}'$. The double degeneracy of the $E'$ modes becomes evident under oblique incidence, which lifts this degeneracy, as will be discussed in Sec.~\ref{sec:AngTune}.

Additionally, modes 1 and 3 are spectrally very close for equilateral configuration; however, they do not couple because they belong to different irreps of the symmetry group. The observed reduction in the scattering intensity of peak II for geometries $20^\circ \leq \theta \leq 30^\circ$ results from the reduced strength of the individual dipoles associated with mode 2. 
    
\subsubsection{\textbf{Oblique incidence for accessing additional modes}}
\label{sec:AngTune}

Under normal incidence with in-plane $s$-polarization, an atomic trimer excites only three fundamental modes, resulting in limited spectral tunability under geometric deformation. Accessing additional modes to increase tunability requires varying the angle of incidence, which excites otherwise dark modes. We therefore focus on the excitation of the remaining in-plane modes under oblique incidence. All modes contributing to the observed spectral features are identified  and their characteristics are examined using eigenmode analysis. This is crucial for selective mode excitation and for gaining control over directionality, scattering strength, and coupling to other emitters. We first discuss the isosceles configuration belonging to the $C_{2v}$ symmetry group, which supports all six in-plane modes. The linear and equilateral configurations, belonging to $D_{\infty h}$ and $D_{3h}$ symmetry groups, respectively, are discussed separately.
    
\paragraph{Isosceles trimer.} \label{para:Iso_IncAng}	
An isosceles trimer of leg length $l=0.07\lambda_a$ is illuminated by an $s$-polarized plane wave incident at an angle $\phi$. To investigate the influence of the incident angle on the optical response, we analyze the scattering for $\phi$ ranging from $\phi=0^\circ$ (normal incidence, with the field parallel to the base) to $\phi=90^\circ$ (field perpendicular to the base). This angular range fully captures the response of the antenna, since an isosceles trimer possesses $C_{2v}$ symmetry, which makes the scattering invariant under reflection about the principal axis. 

\begin{figure}[h!]
    \centering
    \includegraphics[width = 1\textwidth]{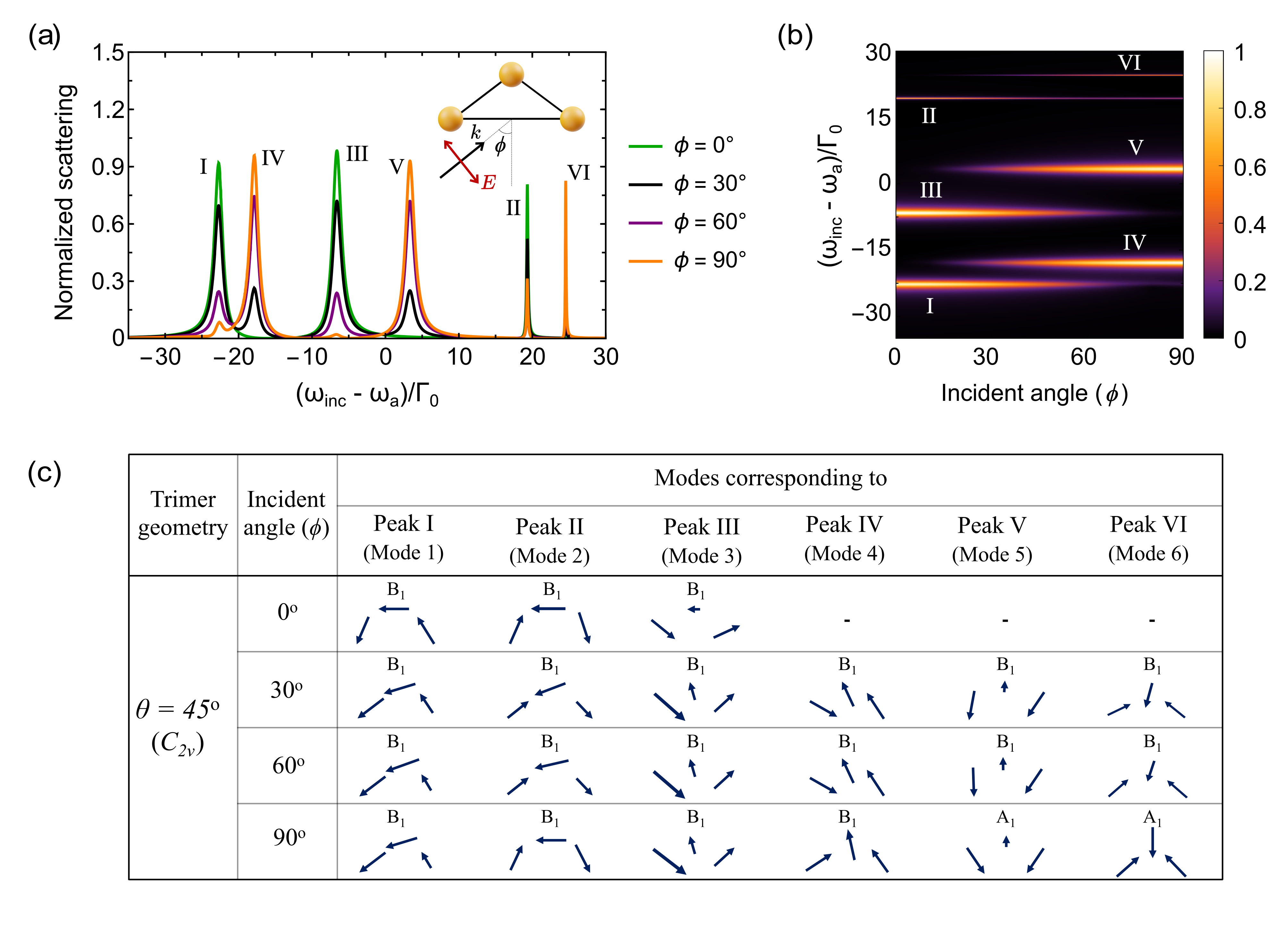}
    \caption{Optical response of the isosceles atomic trimer of leg length $l=0.07\lambda_a$ and base angle $\theta =45^\circ$, under in-plane oblique $s$-polarized incidence, along with mode characteristics. (a) Normalized scattering spectra showing six peaks corresponding to the excitation of six pure in-plane modes, labeled I-VI. An $s$-polarized plane wave is incident at an angle $\phi$ (see inset). (b) Scattering intensity  as a function of $\phi$, illustrating the relative strength of the modes. (c) Mode classification for  $\theta=45^\circ$: all modes belong to the $C_{2v}$ point group and are classified as $\text{B}_1$ and $\text{A}_1$ irreps.}
    \label{fig:iso_incAngle}
\end{figure}

The results are shown in Fig. \ref{fig:iso_incAngle}, which presents the scattering spectra for different incident angles, along with the mode characteristics at a representative base angle $\theta=45^\circ$. Figure \ref{fig:iso_incAngle}(a) shows the normalized scattering spectra for incident angles $\phi=0^\circ, 30^\circ, 60^\circ$, and $90^\circ$. The inset depicts the antenna geometry, with the incident angle and polarization marked. The spectra exhibit six peaks (labeled I - VI), whose intensities depend on the angle of incidence, while their spectral positions remain unchanged as the incident angle varies. Eigenmode analysis reveals that each spectral peak corresponds to a single pure in-plane mode. Figure $\ref{fig:iso_incAngle}$(a) shows that for a base angle of $\theta=45^\circ$, the modes are spectrally well separated. This holds for base angles from $\sim 0^\circ$ to $\sim 50^\circ$ enabling selective excitation of the individual antenna modes for practical purposes. To clearly illustrate the growth and gradual weakening of the modes with incident angle, Fig. \ref{fig:iso_incAngle}(b) shows the scattering as a function of $\phi$. The orientations of the induced dipoles corresponding to each spectral peak, arising from the respective pure modes (labeled 1 - 6), are tabulated in Fig. \ref{fig:iso_incAngle}(c). All modes belong to the $C_{2v}$ symmetry point group and can be classified into the $\text{B}_1$ and $\text{A}_1$ irreps. When the high symmetry of the equilateral configuration is broken and angular incidence is employed, (i) mode 2 ($\text{E}'$) splits into new modes 2 and 5, and (ii) mode 3 ($\text{E}'$) splits into new modes 3 and 4. This degeneracy lifting is further discussed in Sec. \ref{App.C}. Mode 6 is dark under normal incidence and becomes accessible under oblique incidence for base angles $0^\circ \leq \theta \lesssim 55^\circ$.
	
\begin{figure}
\centering
    \includegraphics[width = 1\textwidth]{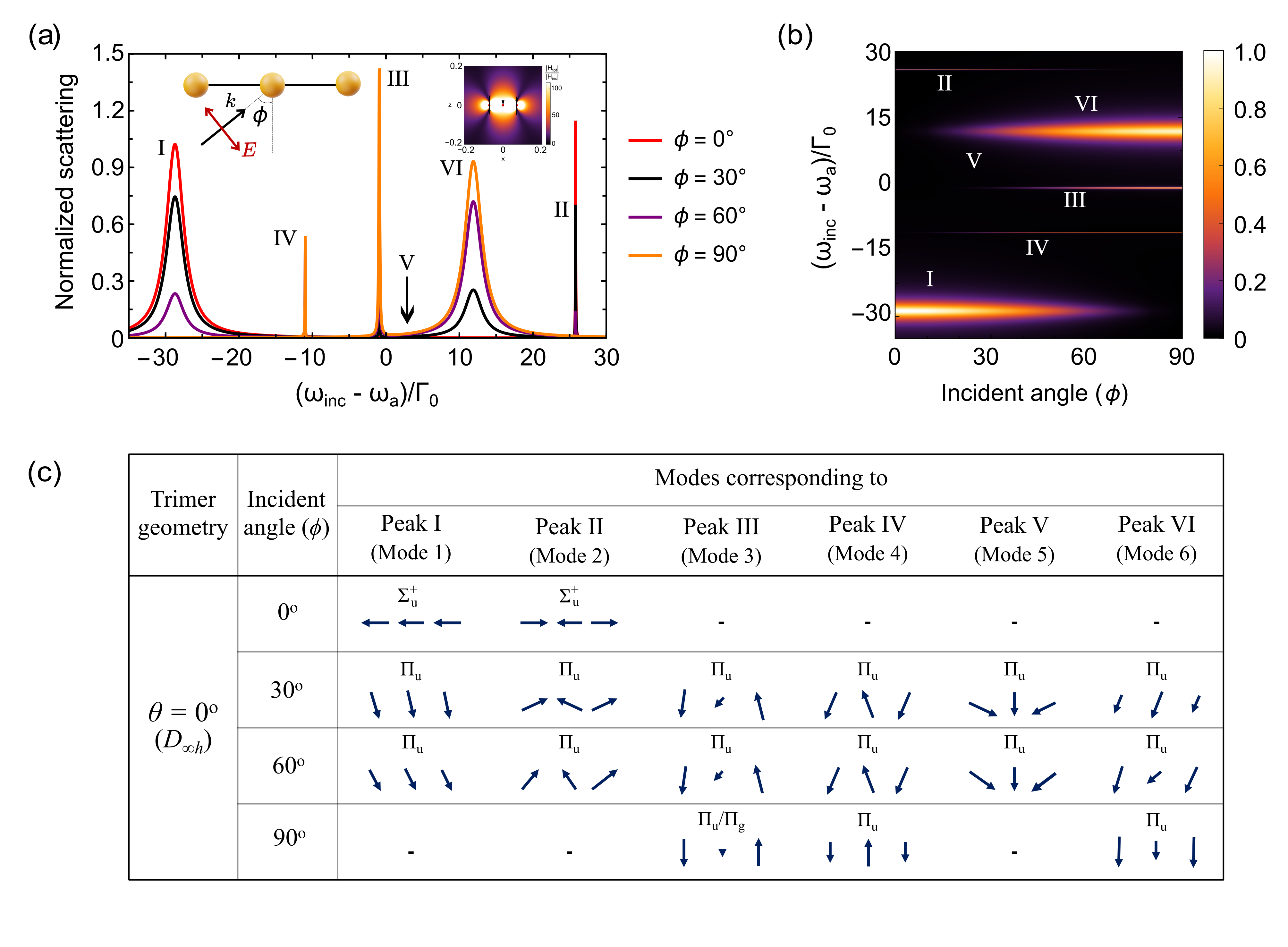}
    \caption{Optical response of the linear atomic trimer with interatomic separation $l=0.07\lambda_a$, for in-plane oblique $s$-polarized incidence, along with mode characteristics. 
    (a) Normalized scattering spectra showing five peaks (I-IV and VI) corresponding to the excitation of five in-plane modes, and an additional small peak (V) indicated by an arrow. Magnetic field enhancement at peak III is shown to highlight its magnetic nature. 
    (b) Scattering intensity showing the relative strength of the modes as a function of the incident angle $\phi$. 
    (c) Mode classification for linear trimer: all modes belong to $D_{\infty h}$ point group and are classified as: $\Sigma_u^+$ irrep under normal incidence, and $\Pi_u$ irreps under oblique incidence.} 
    \label{fig:Lin_incAngle}
\end{figure}

\paragraph{Linear trimer.} \label{para:linear_IncAng}
Next, a linear trimer with interatomic separation  $l=0.07\lambda_a$ is investigated under the same conditions as before: illumination by an $s$-polarized plane wave. The normalized scattering shown in Fig. \ref{fig:Lin_incAngle}(a) exhibits five distinct peaks (labeled I-IV and VI) and a weak peak (labeled V, indicated by the arrow), as the incident angle varies from $\phi=0^\circ$ (normal incidence, electric field parallel to the trimer chain) to $\phi=90^\circ$ (electric field perpendicular to the trimer chain). This angular range is necessary to fully capture the antenna response, since the linear trimer possesses $D_{\infty h}$ symmetry. Eigenmode analysis reveals that each spectral peak corresponds to a single pure in-plane mode. The intensities of the peaks depend on the angle of incidence, while their spectral positions remain unchanged as the incident angle varies. Figure \ref{fig:Lin_incAngle}(b) provides a clear visualization of the growth and gradual weakening of the spectral peaks as the incident angle changes. The evolution of the orientations of the dipoles is obtained as before and is tabulated in Fig. $\ref{fig:Lin_incAngle}$(c). The modes corresponding to each peak are identified, and their irreps are indicated. All modes belong to the $D_{\infty h}$ symmetry point group and can be classified into: $\Sigma_u^+$ irrep under normal incidence and $\Pi_u$ irrep under oblique incidence. 
\\
The spectral positions of peaks II and VI are interchanged relative to their positions in the isosceles geometry. This indicates a crossing of the modes 2 and 6, corresponding to the peaks II and VI respectively, as the linear trimer is tuned toward the equilateral configuration. This mode crossing is discussed for an angle of incidence $\phi = 30^\circ$ in Sec. \ref{App.C}, where it occurs at a base angle $\theta \approx 37^\circ$. Since both modes belong to the same irrep, they interact at the crossing, resulting in a dipole orientation that arises from their mixing. 

Mode 3, corresponding to peak III, belongs to the $\Pi_{u}$ irrep. As $\phi \rightarrow 90^\circ$, the strength of the central dipole becomes negligible (see Fig. \ref{fig:Lin_incAngle}(c)), and in this limit the mode's irrep is $\Pi_g$. Owing to the antiparallel orientation of the two remaining dipoles, mode 3 exhibits a strong magnetic character, with its intensity reaching a maximum at $\phi=90^\circ$. The normalized total magnetic field distribution $|\mathbf{H}_\text{tot}|/|\mathbf{H}_\text{inc}|$ (where $|\mathbf{H}_\text{tot}|$ and $|\mathbf{H}_\text{inc}|$ are the total and incident magnetic fields, respectively) at $\phi=90^\circ$ is shown in the inset of Fig. \ref{fig:Lin_incAngle}(a), demonstrating the enhancement of magnetic field. This mode is analogous to the antisymmetric (magnetic) mode of an atomic dimer discussed in Ref.~\cite{alaee2020quantum}.

\begin{figure} [h!]
    \centering
    \includegraphics[width = 0.7\textwidth]{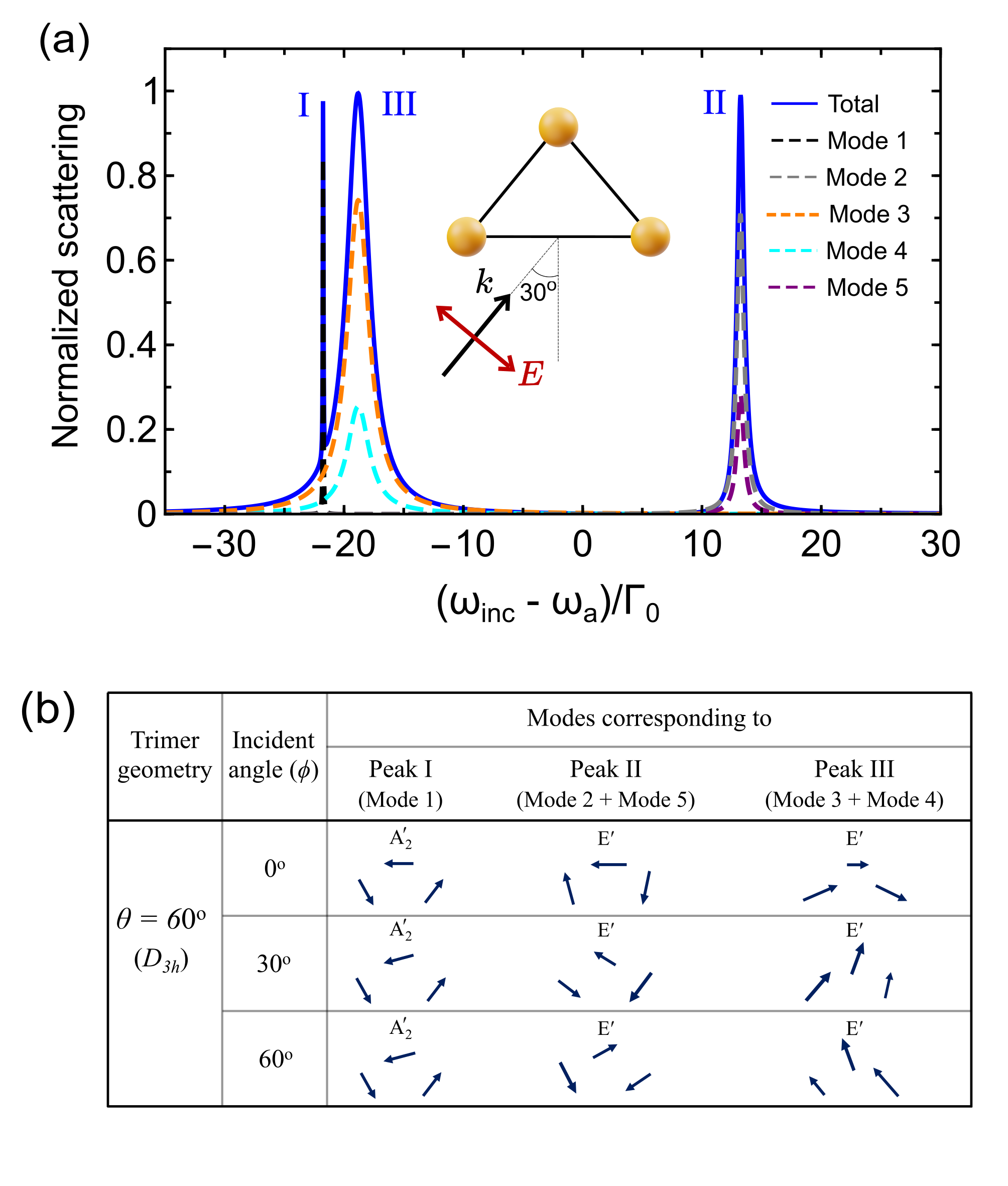}
        \caption{Optical response of an equilateral trimer with interatomic separation $l=0.07\lambda_a$, under in-plane oblique $s$-polarized incidence, along with mode characteristics. 
        (a) Normalized scattering spectra, showing three peaks arising from five in-plane modes. Peak I arises exclusively from mode 1, while peaks II and III originate from modes 2 and 5, and modes 3 and 4, respectively. 
        (b) Mode classification for equilateral trimer: all modes belong to the $D_{3h}$ point group, and are classified as $\text{A}^{'}_{2}$ and $\text{E}^{'}$ irreps.} 
    \label{fig:Equi_incAngle}
\end{figure}

\paragraph{Equilateral trimer.} \label{para:Equi_IncAng}
We finally investigate the equilateral trimer with interatomic separation  $l=0.07\lambda_a$ under oblique incidence by an $s$-polarized plane wave at an angle~$\phi$. The total spectrum results from the excitation of five fundamental in-plane modes, whose strengths depend on the incident angle, while their spectral positions are set only by the antenna size. Fig. \ref{fig:Equi_incAngle}(a) shows the normalized total scattering along with the modal contributions to the three observed spectral peaks for $\phi = 30^\circ$. Peak I is a pure mode (mode 1). Peaks II and III results from degenerate pairs: the former from modes 2 and 5, the latter from modes 3 and 4. The modal contributions to the three spectral peaks vary with incident angle, with peak I corresponding to a single mode and peaks II and III arising from two-mode superpositions. However, the total scattering intensity, summed over all peaks, remains constant due to high symmetry of the equilateral antenna ($D_{3h}$), rendering the overall scattering spectrum invariant with respect to incident angle.
Further, the orientations of individual dipole moments corresponding to each peak are tabulated in Fig. \ref{fig:Equi_incAngle}(b) along with their respective irreps for $\phi = 0^\circ,30^\circ$, and $60^\circ$. 
For the equilateral geometry ($D_{3h}$ point group), incident angles $0^\circ-60^\circ$ are sufficient to fully map the antenna response due to its threefold rotational symmetry about the principal axis, which makes the response periodic with $60^\circ$.
	
Mode 1 (peak I) belongs to the $\text{A}^{'}_{2}$ irrep and exhibits strong magnetic character. This behavior persists for all incident angles, with total forward scattering along the incidence direction, owing to the high symmetry ($D_{3h}$) of the equilateral trimer. The high symmetry preserves the strength of the effective magnetic moment, which, together with the electric moment of mode 3, satisfies the Kerker condition \cite{alaee2020kerker}. All other modes belong to the $\text{E}^{'}$ irrep. Although peaks I and III are spectrally close, their modes do not couple because they belong to different irreps. 

\begin{figure}
\centering
    \includegraphics[width = 0.6\textwidth]{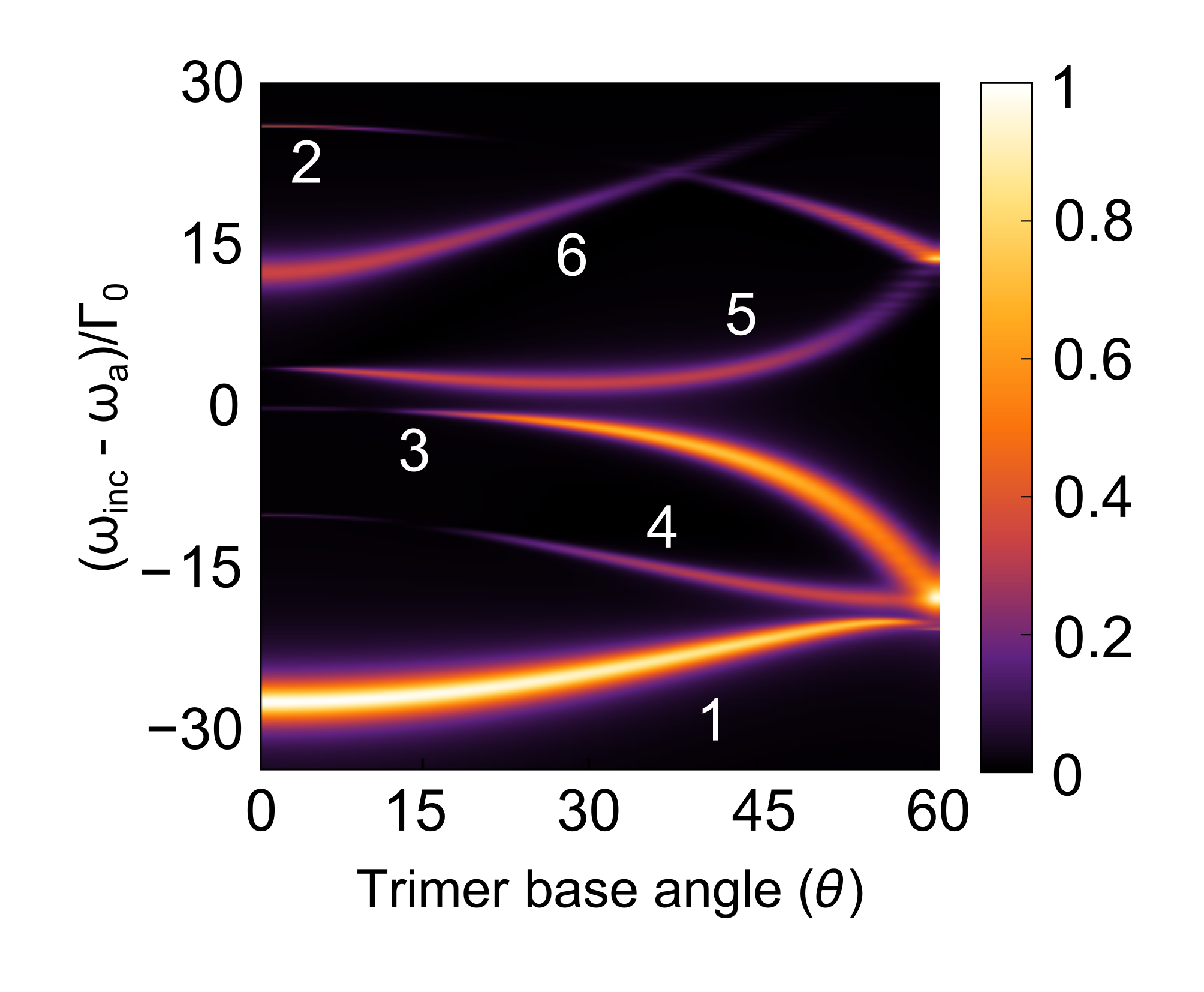}
    \caption{Map of the scattering cross section showing the evolution of the excited modes (labeled 1–6) as the trimer transforms from linear to equilateral configuration, as a function of $\theta$, under in-plane $s$-polarized incidence. As the equilateral symmetry is broken, the degeneracies of the $\text{E}'$ modes are lifted: the first pair evolves into modes 2 and 5, and the second pair evolves into modes 3 and 4. Also, modes 2 and 6 cross at base angle $\theta \approx 37^\circ$. The leg length is fixed at $l = 0.07 \lambda_{a}$, and the incident angle $\phi = 30^\circ$.}
\label{fig:Demo_psi=30}
\end{figure}

\paragraph{Modal degeneracy-lifting and spectral crossing.} 
\label{App.C}
To better understand the modal evolution with trimer geometry under in-plane $s$-polarized plane wave excitation, we map the scattering cross section as a function of detuning while gradually varying the base angle $\theta$ from $0^\circ$ to $60^\circ$. The resulting map is shown in Fig. \ref{fig:Demo_psi=30} for a fixed incident angle $\phi = 30^\circ$, where the spectral shifts are clearly visible. These results provide a clear visualization of the degeneracy lifting upon symmetry breaking and the spectral crossing during geometric tuning, which were mentioned in Secs. \ref{para:Iso_IncAng} and \ref{para:linear_IncAng}, respectively. The six modes (labeled 1-6) exhibit distinct spectral evolution with $\theta$, with degeneracies emerging as the equilateral configuration is approached. The equilateral trimer belongs to the high-symmetry point group $D_{3h}$, under which specific modes form doubly degenerate E$'$ pairs, as discussed in Sec. \ref{para:Equi_IncAng}. Upon breaking the equilateral symmetry, these degeneracies are lifted, causing peaks II and III at $\theta=60^\circ$ to split into modes 2 and 5, and modes 3 and 4, respectively. This degeneracy lifting due to symmetry breaking is evident in Fig. \ref{fig:Demo_psi=30}.

The scattering intensity of the individual modes varies significantly with the base angle $\theta$, indicating that the geometric tuning affects not only the resonance frequencies but also the strength with which the modes are excited. These intensity variations arise from the geometry-dependent coupling between the incident field and the eigenmodes. As $\theta$ changes, the relative orientations and amplitudes of the atomic dipoles within each eigenmode evolve, modifying both the effective dipole moment and its projection onto the incident electric field. In particular, the scattering intensities of modes 2 and 4 reduce in the ranges $20^\circ \leq \theta \leq 30^\circ$ and around $15^\circ$, respectively. In addition, the intensity of mode 6 decreases with $\theta$ and vanishes beyond $\theta \approx 55^\circ$.

A crossing between modes 2 and 6 is observed during the geometric tuning at $\theta \approx 37^\circ$, as shown in Fig. \ref{fig:Demo_psi=30}. Since both modes belong to the same irrep, they interact at the crossing, resulting in a dipole orientation due to their mixing. 

\subsection{Out-of-plane collective modes}
\label{subsec:s_polarization}

We now focus on the out-of-plane modes of the antenna excited under in-plane $p$-polarized incidence, completing the discussion of all nine eigenmodes of the trimer. For $p$-polarization, the induced electric dipole moments are normal to the trimer plane, and the resulting modes are therefore referred to as out-of-plane modes. The schematic of the system is shown in Fig. \ref{fig:Trimer_tuning}, where the interatomic distances in the linear configuration and the leg lengths in the isosceles configuration are both maintained at $l = 0.07 \lambda_{a}$, and $E_p$ denotes the incident polarization. For $p$-polarization, the modified multipole moments, $p'^{y}_\text{eff}$, $m'^{x}_\text{eff}$ and $m'^{z}_\text{eff}$, are obtained by incorporating the effects of the shifted scattering centers into Eqs. (\ref{eq:pyeff}), (\ref{eq:mxeff}), and (\ref{eq:mzeff}), and are given below
\begin{equation} \label{eq:pyeffp}
    \begin{split}
        p'^{y}_\text{eff} = \sum\limits_{i=1}^{3} \Bigg[ p^{y}_{i} \, j_{0} \!\left(k|\mathbf{r} - \mathbf{d}_{e}|_{i}\right) - \frac{k^{2}}{2} p^{y}_{i} (r^{x} - d^{x}_{e})_i^{2} \, \frac{j_{2}\!\left(k |\mathbf{r} - \mathbf{d}_{e}|_{i}\right)}{(k|\mathbf{r}-\mathbf{d}_{e}|_{i})^{2}} \Bigg],
\end{split}
\end{equation}
\begin{equation} \label{eq:mxeffp}
    m'^{x}_\text{eff} = \frac{3 i \omega}{2} \sum\limits_{i=1}^{3} \Bigg[ (p^{y}_{i} (r^{z}-d_{m}^{z})_{i}) \, \frac{j_{1}\!\left(k|\mathbf{r}-\mathbf{d}_{m}|_{i}\right)}{k |\mathbf{r}-\mathbf{d}_{m}|_{i}} \Bigg],
\end{equation}
\begin{equation} \label{eq:mzeffp}
    m'^{z}_\text{eff} = -\frac{3 i \omega}{2} \sum\limits_{i=1}^{3} \Bigg[ (p^{y}_{i} (r^{x}-d_{m}^{x})_{i}) \, \frac{j_{1}\!\left(k|\mathbf{r}-\mathbf{d}_{m}|_{i}\right)}{k |\mathbf{r}-\mathbf{d}_{m}|_{i}} \Bigg], 
\end{equation}
	
\noindent	
where $d_{e}^{x}, d_{e}^{z}, d_{m}^{x}$, and $d_{m}^{z}$ represent the shifts of the moments from the geometric center along the respective directions. 
These modified moments $p'^{y}_\text{eff}$, $m'^{x}_\text{eff}$ and $m'^{z}_\text{eff}$ are substituted into Eq. (\ref{eq:MPD}) for the calculation of the scattering cross section.

\begin{figure}[h!]
    \centering
    \includegraphics[width = 0.5\textwidth,trim = 1pt 0pt 0pt 0pt, clip]{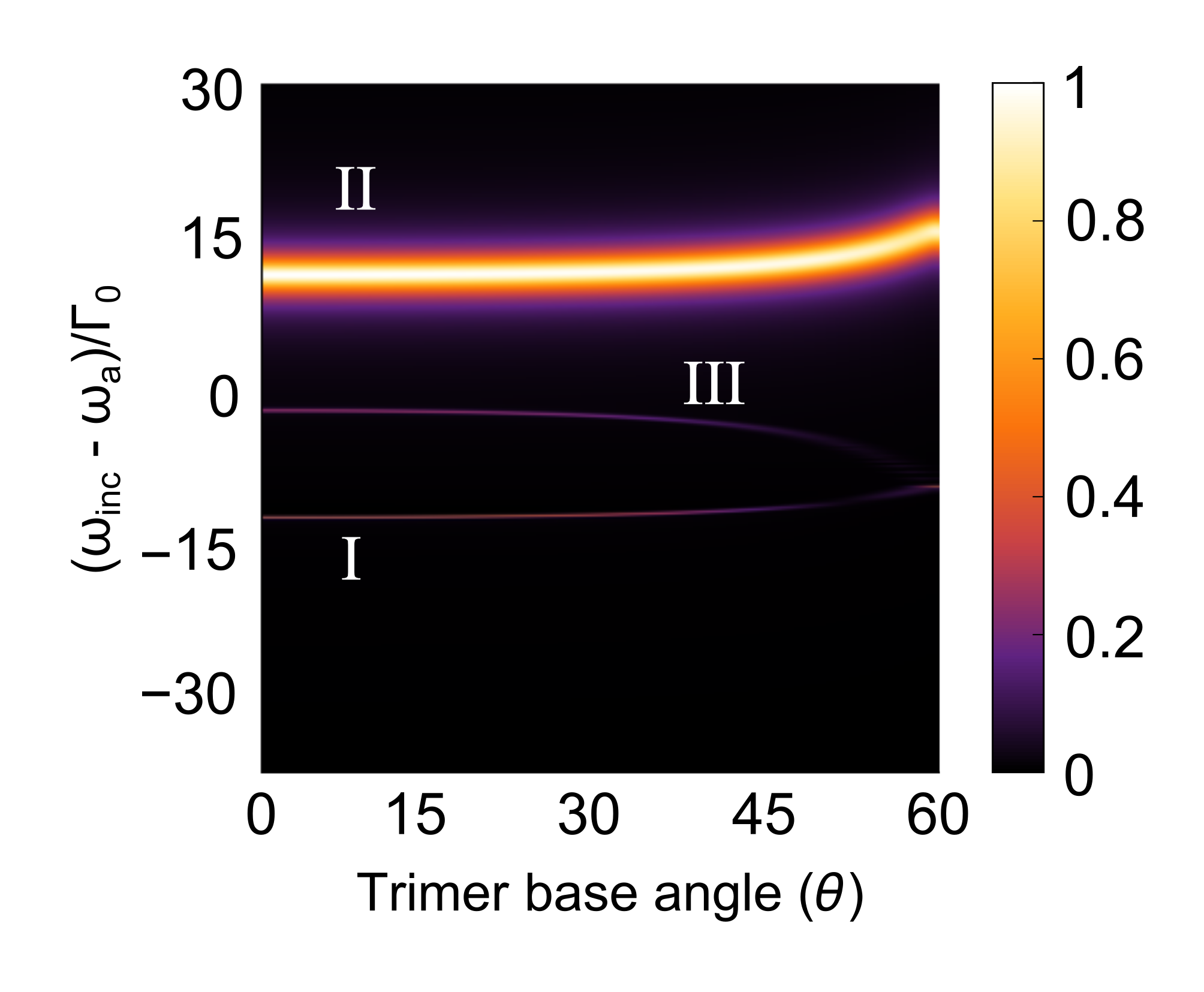}
    \caption{Geometric tuning of the optical response of a trimer under in-plane $p$-polarized illumination. Scattering cross section showing the evolution of the spectral peaks (labeled I - III) as the trimer is transformed from linear to equilateral configuration as a function of $\theta$, for $\phi=30^\circ$ and $l = 0.07 \lambda_{a}$. }
    \label{fig:trimer_scat_s_pol}
\end{figure}

Total response of the antenna is studied for configurations ranging from linear to equilateral geometry by calculating the scattering cross section as a function of detuning. The result is shown in Fig. \ref{fig:trimer_scat_s_pol}, for an incident angle $\phi=30^\circ$, where the evolution of scattering peaks I, II, and III are clearly seen. Two spectrally well-separated modes (modes 7 and 8, corresponding to peaks I and II) are excited for normal incidence, while mode 9 (peak III) appears under oblique incidence. Modes 7 and 9 are degenerate for an equilateral configuration. The spectral shift under geometric tuning is much smaller for $p$-polarized incidence than for $s$-polarization, and this behavior holds for all angles of incidence. 

\begin{table}[h]
    \centering
    \caption{Mode classification for linear, isosceles and equilateral geometries under oblique in-plane $p$-polarized incidence: all modes contributing to the three spectral peaks I, II, and III (modes 7, 8, and 9, respectively) are categorized according to their corresponding symmetry point groups and irreps. The green arrow denotes the induced magnetic dipole arising from the associated circular current distribution.}
    \includegraphics[width = 1\textwidth,trim = 1pt 0pt 0pt 0pt, clip]{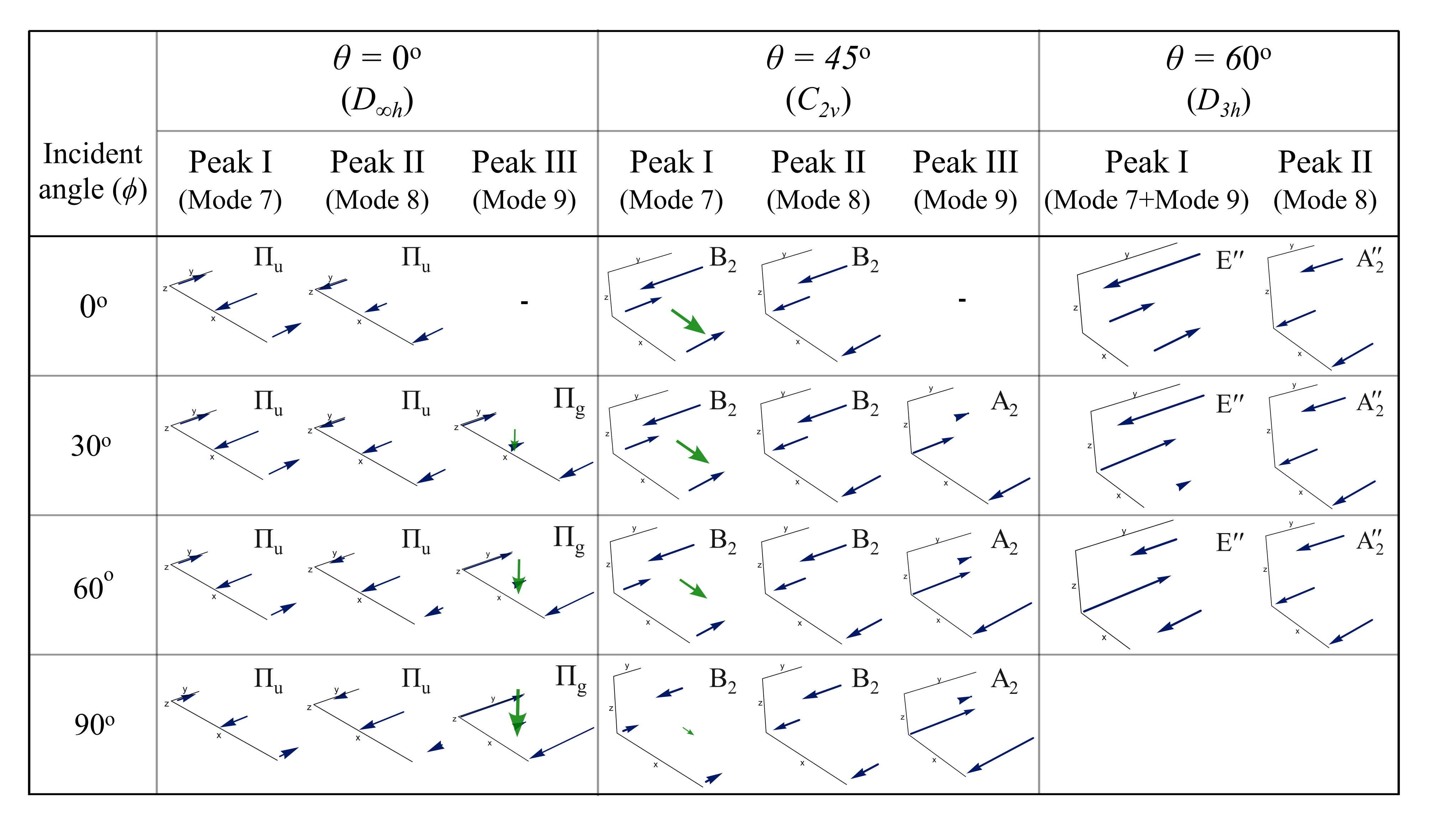}
    \label{tab:trimer_scat_s_pol}
\end{table}

Further, we calculate the strength and orientation of the individual dipoles for $D_{\infty h}$, $C_{2v}$, and $D_{3h}$ geometries and examine how the collective modes evolve as the antenna configuration is tuned. 
The results are summarized in Table \ref{tab:trimer_scat_s_pol}, along with the corresponding symmetry point groups and associated irreps.
Geometric tuning of the trimer from a linear to an equilateral configuration under $p$-polarization shows that mode 7 evolves as $\Pi_u \rightarrow \text{B}_2 \rightarrow \text{E}''$, while mode 8 evolves as $\Pi_u \rightarrow \text{B}_2 \rightarrow \text{A}_2''$. In contrast, mode 9 evolves as $\Pi_g \rightarrow \text{A}_2 \rightarrow \text{E}''$. 
The arrows representing the dipoles illustrate the nature and symmetry of the modes, while their lengths qualitatively indicate the relative strengths. 
When the linear geometry is progressively distorted, mode 7 develops an effective magnetic dipole moment along the $x$-direction. 
Mode 9 also exhibits a strong magnetic character due to its circular current configuration, with an effective magnetic dipole moment directed along the $z$-direction. 
Both induced magnetic dipoles are indicated in the table by green arrows. 
The magnetic nature of mode 9 is further explored in Sec. \ref{sec:emission}. 

\subsection{Out-of-plane incidence}

\begin{figure}[h!]
    \centering
    \includegraphics[width = 1\textwidth,trim = 1pt 0pt 0pt 5pt, clip]{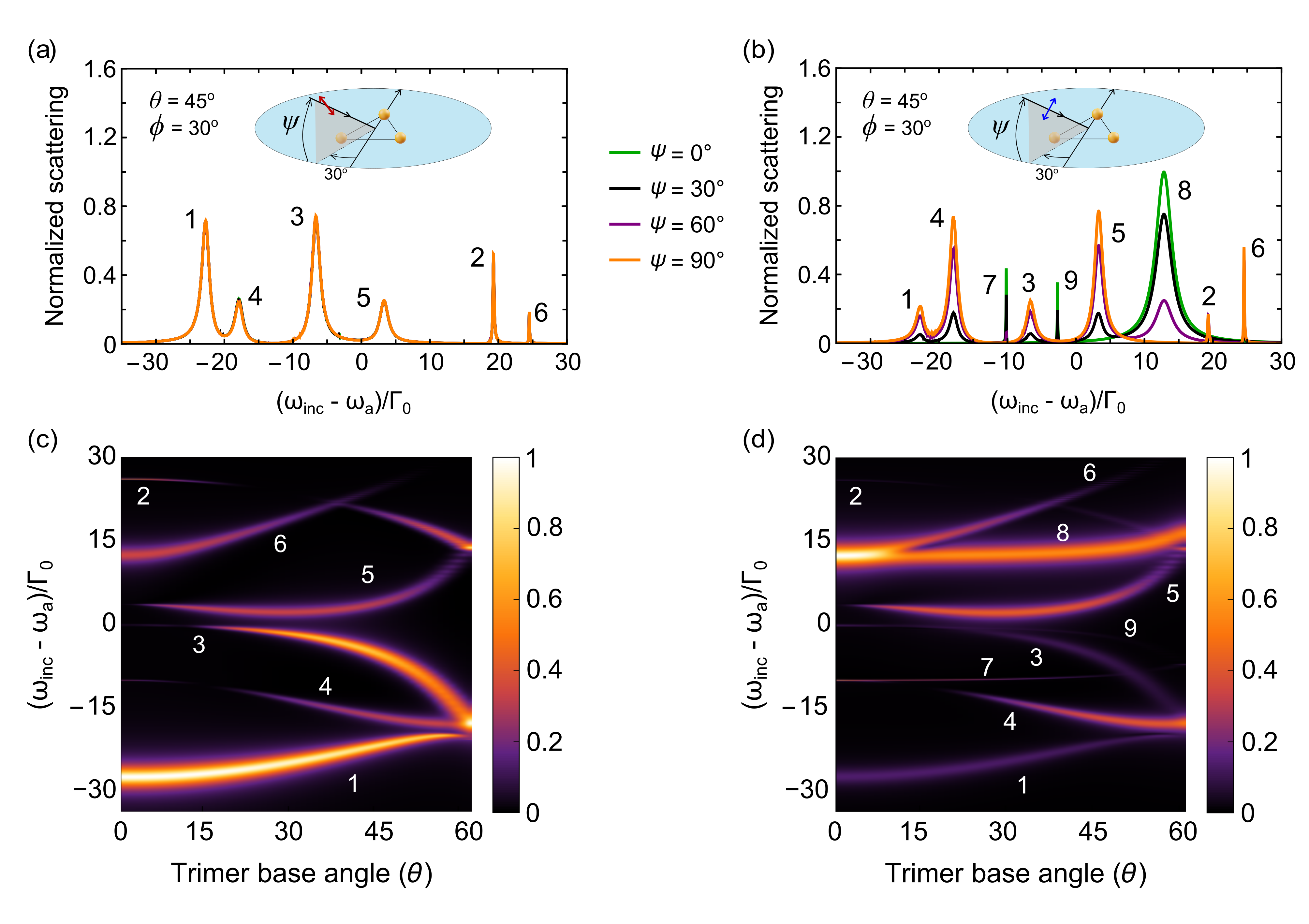}
    \caption{Optical response of an atomic trimer under out-of-plane incidence, with $\psi$ denoting the out-of-plane angle, for a fixed in-plane angle $\phi=30^\circ$. (a, b) Normalized scattering spectra as a function of detuning for an isosceles configuration with $\theta=45^\circ$ at $\psi=0^\circ, 30^\circ, 60^\circ,$ and $90^\circ$, under $s$- and $p$-polarized plane wave excitation, respectively. (c, d) Maps of the scattering cross section showing the evolution of the excited modes with base angle $\theta$ as the trimer transforms from linear to equilateral configuration, under $s$- and $p$-polarized excitation, respectively. Insets show the corresponding incident light configuration. The interatomic distance for the equilateral and linear trimers and the leg length for the isosceles trimer are fixed at $l=0.07\lambda_{a}$.}
    \label{fig:Elevated_inc}
\end{figure}

Having established the excitation of all nine eigenmodes under in-plane incidence, with the six in-plane modes (1-6) excited by $s$-polarized light and the three out-of-plane modes (7-9) by $p$-polarized light, we now turn to out-of-plane incidence. The dependence of the spectra on the out-of-plane angle $\psi$ is examined for the isosceles, equilateral, and linear geometries (Fig. \ref{fig:Trimer_schematic}). Although the introduction of the out-of-plane angle $\psi$ does not modify the set of accessible eigenmodes, it redistributes their modal amplitudes. In addition, variation of the incident direction and polarization with $\psi$ reorients the excited dipole configurations, while preserving the underlying modal character. Consequently, the response under out-of-plane incidence can be understood within the modal framework  established for in-plane incidence. 

For in-plane incidence at $\phi=0^\circ$, as discussed in Secs. \ref{subsubsec:NormInc} and \ref{subsec:s_polarization}, not all modes are excited, and a finite in-plane angle $\phi$ is required to access the complete set of eigenmodes. Therefore, to investigate the spectral evolution of the complete response under out-of-plane incidence, we fix the in-plane angle at $\phi=30^\circ$. Figure \ref{fig:Elevated_inc} shows the dependence of the trimer response on the out-of-plane angle $\psi$ for different configurations.
For $s$-polarized incidence, varying $\psi$ preserves the same set of in-plane modes (1-6). In contrast, for $p$-polarized incidence, the polarization vector progressively tilts out of the trimer plane as $\psi$ increases, enabling simultaneous excitation of both in-plane (1-6) and out-of-plane (7-9) modes. The evolution of these modes with the trimer geometry further reveals the degeneracies associated with the high-symmetry linear and equilateral configurations. As these symmetries are broken, the degeneracies are lifted, leading to the splitting and crossing of the modes. Such degeneracies are particularly interesting because they can occur independently among in-plane modes and among out-of plane modes, as well as between in-plane and out-of-plane modes.

For out-of-plane incidence, the modified multipole moments for $s$- and $p$-polarizations are obtained from Eqs. (\ref{eq:pxeffp}) - (\ref{eq:mzeffp}), which are then substituted into Eq. (\ref{eq:MPD}) for the calculation of the scattering cross section. Figures \ref{fig:Elevated_inc} (a) and (b) show the scattering spectra for an isosceles configuration ($\theta=45^\circ$) at $\psi = 0^\circ, 30^\circ, 60^\circ,$ and $90^\circ$, for $s$- and $p$-polarizations, respectively. This angular range $0^\circ \leq \psi \leq 90^\circ$ fully captures the antenna response, since the scattering remains symmetric about $\psi=90^\circ$. The mode numbering (1-9) follows that established for in-plane incidence, enabling direct identification of the excited eigenmodes associated with each spectral peak. For $s$-polarization, the scattering spectrum exhibits six peaks corresponding to the six in-plane modes and remains essentially invariant with increasing $\psi$ [Fig.~\ref{fig:Elevated_inc}(a)]. Although varying $\psi$ modifies the phase of the incident field at the three atoms and thus its projection onto the individual eigenmodes, the corresponding quantities $\left|\mathbf m_n^*\cdot\mathbf E_\text{inc}\right|^2$ vary only weakly. After accounting for the mutual coupling between the eigenmodes, the corresponding modal coefficients $b_n$ remain roughly constant (with variations only in the second or third decimal places), as quantified by $|b_n|^2$. Consequently, the excitation strengths of the six modes, and hence the spectral peak intensities remain effectively invariant with $\psi$. Under $p$-polarized incidence, the tilted polarization vector acquires both in-plane and out-of-plane components, thereby enabling the excitation of in-plane modes (1-6) in addition to the out-of-plane modes (7-9) [Fig.~\ref{fig:Elevated_inc}(b)]. As $\psi$ increases from zero, the spectral intensities of the out-of-plane modes gradually decrease, whereas those of the in-plane modes increase. This redistribution follows from the decrease in the out-of-plane component and increase in the in-plane component of the incident field as $\psi$ increases. At $\psi = 90^\circ$, the polarization vector lies entirely within the trimer plane, and consequently only the in-plane modes remain.

When the trimer  attains linear and equilateral configurations, the spectra become relatively simple owing to the degeneracy of a few pairs of modes. Since the evolution of the spectra with $\psi$ is evident from the isosceles results, to discuss the spectral changes as the equilateral and linear configurations are reached, we map the scattering as a function of the base angle $\theta$ for a fixed $\psi=45^\circ$. This also reveals the mode degeneracies as the high symmetry equilateral and linear configurations are approached. To visualize these features, in Figs. \ref{fig:Elevated_inc}(c) and (d) we present the total scattering as a function of the base angle $\theta$ at $\phi=30^\circ$ and $\psi=45^\circ$, for both $s$- and $p$-polarizations, respectively. The shifts of the individual spectral peaks with $\theta$ are consistent with those observed under in-plane incidence. For $s$-polarization, all six in-plane modes are excited, and their spectral evolution is shown in Fig. \ref{fig:Elevated_inc}(c). These six modes remain excited throughout the evolution towards the linear configuration. At the equilateral configuration, modes 2 and 5, as well as modes 3 and 4 become degenerate. Furthermore, modes 2 and 6 cross at $\theta\approx37^\circ$, while mode 6 vanishes as the equilateral configuration is attained. The intensity variations of modes 2 and 4 follow a trend similar to that observed for in-plane incidence. For $p$-polarization, all nine modes are excited, and their spectral evolution, shown in Fig. \ref{fig:Elevated_inc}(d), is richer. At the linear configuration, three sets of in-plane and out-of-plane modes become degenerate, whereas at the equilateral configuration, degeneracy occurs separately within the in-plane and out-of-plane modes. Particularly, modes 3 and 9, 4 and 7, and 6 and 8 become degenerate as the linear configuration is approached, reducing the total number of spectral peaks to six. In contrast, at the equilateral configuration, modes 2 and 5, and modes 3 and 4 become degenerate among the in-plane modes, while modes 7 and 9 become degenerate among the out-of-plane modes. In addition, mode 6 vanishes, further reducing the total number of spectral peaks to five. 

The spectral evolution with $\psi$ for the equilateral and linear geometries follows the same trends as observed for the isosceles geometry and can therefore be inferred from Figs. \ref{fig:Elevated_inc}(a) and (b). For $s$-polarized incidence, the intensity of mode 1 in the equilateral configuration is maximal for in-plane incidence, owing to its circulating dipole configuration. As the incidence becomes increasingly out-of-plane, the circulating character of the mode, and consequently the magnetic character of the antenna, decreases and vanishes at $\psi=90^\circ$. At the linear configuration, mode 3 follows a similar evolution with increasing $\psi$ and vanishes at $\psi=90^\circ$. 

\section{\MakeUppercase{Detuning-Controlled Forward-Backward Scattering Switching}} \label{sec:FB}
	
\begin{figure}[h!]
\centering
\includegraphics[width = 0.8\textwidth]{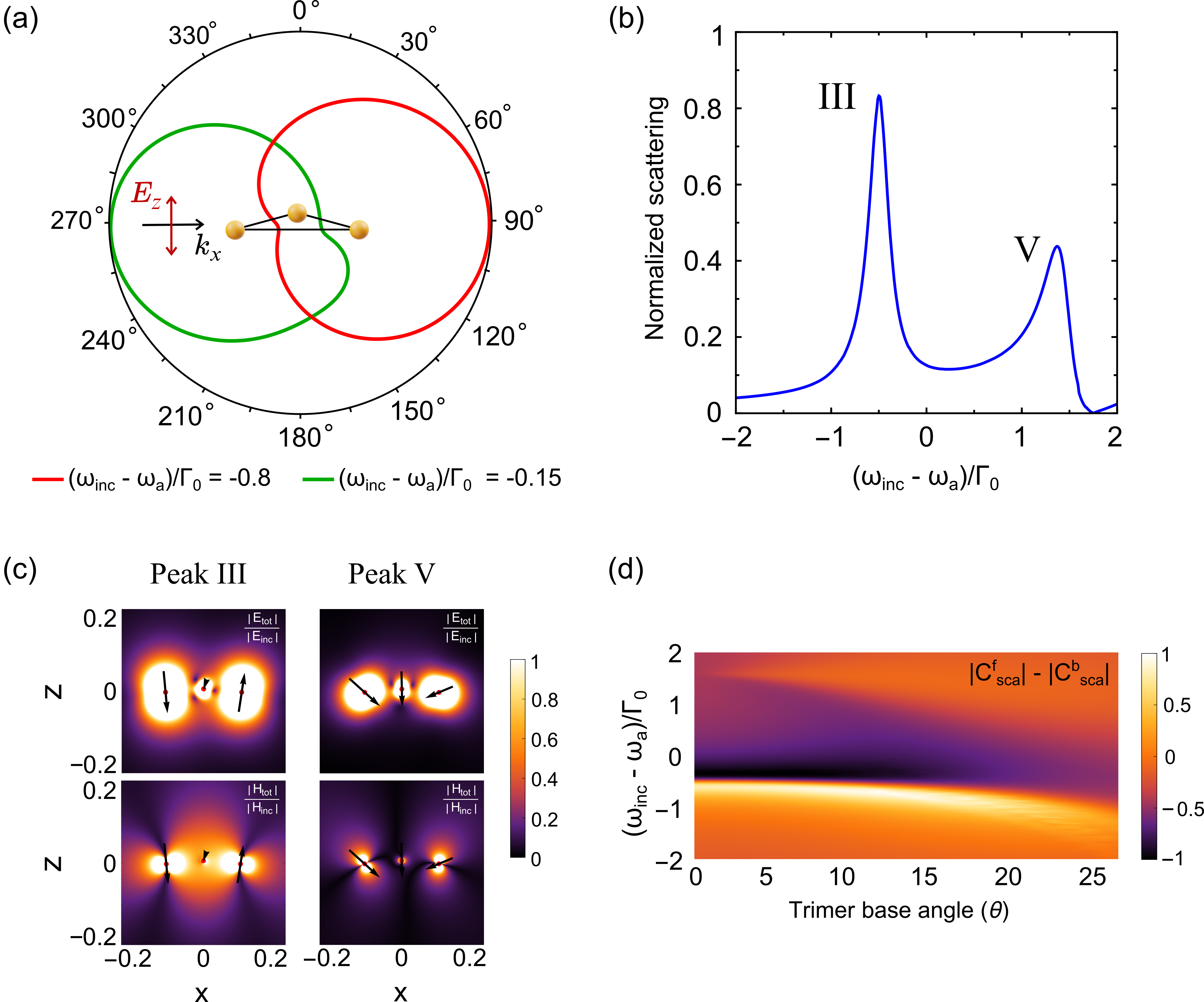}
    \caption{Detuning-driven switching of scattering between forward and backward directions. (a) Radiation patterns showing enhanced forward scattering (red) at frequency detuning $\Delta = -0.8$ and backward scattering (green) at $\Delta = -0.15$, for $l=0.09\lambda_{a}$, $\theta=5^\circ$ under in-plane $s$-polarized plane wave excitation at $\phi=90^\circ$. (b) Normalized scattering cross section as a function of frequency detuning, with peaks III and V marked. (c) Normalized total electric ($|\mathbf{E}_{\text{tot}}|/|\mathbf{E}_{\text{inc}}|$) and magnetic ($|\mathbf{H}_{\text{tot}}|/|\mathbf{H}_{\text{inc}}|$) field distributions at peak III and  V, respectively, revealing the electric and magnetic nature of the corresponding modes. (d) Difference between normalized forward and backward scattering cross sections ($|C_\text{sca}^{f}|$ - $|C_\text{sca}^{b}|$) showing the geometric range $\theta = 0^\circ - 15^\circ$ over which switching is efficient.} 
\label{fig:Backscat_5}
\end{figure}

By exploring collective modes, we demonstrate that detuning enables reversible switching between forward and backward scattering in a nearly linear trimer. Modes 3 and 5, excited under in-plane $s$-polarized incidence, are particularly relevant for controlling directional scattering. Mode 3 has magnetic character and mode 5 electric character; both belong to the same irrep and are spectrally close, favoring interference and thus enabling control over the scattering direction. Switching can be achieved by slightly detuning the incident frequency. However, for a perfectly linear configuration with $l=0.07 \lambda_{a}$, mode 5 is negligible (Fig. \ref{fig:Lin_incAngle}), so the desired control is not achieved. A small increase in $\theta$ enhances mode 5; but the spectral peaks remain well separated, preventing interference. We tune the interatomic separation $l$ to engineer directional switching. Decreasing $l$ further separates the spectral peaks, whereas increasing $l$ brings them closer and enables mode interaction. To clearly resolve the switching behavior while maintaining sufficient mode interaction, we fix $l = 0.09\lambda_a$ for the present analysis. Switching then occurs for nearly linear configurations up to $\theta \approx 15^\circ$ under $s$-polarized incidence with $\phi \approx 90^\circ$ for maximum intensity. Figure \ref{fig:Backscat_5}(a) illustrates this for $\theta = 5^\circ$, where the enhanced total scattering switches from forward to backward at  $\Delta = -0.8$ and $\Delta = -0.15$, respectively, with $\Delta$ denoting the normalized detuning $(\omega_{\mathrm{inc}}-\omega_a)/\Gamma_0$.
 	
To gain better control over scattering switching, we examine strengths of the collective modes and the orientations of the individual electric dipoles. For nearly linear configurations, modes 2 and 5 become comparable in magnitude, as shown in Fig.~\ref{fig:Backscat_5}(b) for $\theta = 5^\circ$, and the orientations of the individual electric dipoles together with the corresponding field maps are shown in Fig.~\ref{fig:Backscat_5}(c). The electric and magnetic character of the modes is evident from these field distributions and from the enhancements of the electric ($|\mathbf{E}_{\text{tot}}|/|\mathbf{E}_{\text{inc}}|$) and magnetic ($|\mathbf{H}_\text{tot}|/|\mathbf{H}_\text{inc}|$) fields at both peaks. Radiation from these dipoles interferes in the far field, giving rise to the observed scattering patterns. The geometric range for effective switching is determined from the difference between the normalized forward and backward scattering cross sections, as shown in Fig. \ref{fig:Backscat_5}(d). As $\theta$ increases, the magnetic character of mode 3 decreases due to the evolution of its dipole moments. Consequently, directional switching diminishes for $\theta \gtrsim 15^{\circ}$.

These results demonstrate that the scattering direction can be reversibly switched from forward to backward solely by tuning the incident frequency, without any geometric modification of the trimer. This establishes that the radiation direction can be dynamically controlled within the same physical system through spectral tuning alone. In contrast to conventional nanoantennas, where reconfigurable radiation patterns typically require geometric symmetry breaking or structural redesign, the present atomic antenna enables spectral control through tuning of the incident frequency. This inherent tunability makes atomic antenna systems experimentally promising for applications such as optical switching and ultrasensitive detection.
	
The frequency precision required to achieve detuning control at $\Delta \sim 0.1$ depends on the specific atom and transition considered. Prominent elements for atomic antennas include Rb, Cs, Yb, among others \cite{rui2020subradiant, endres2016atom, gross2017quantum}. For the $^{85}$Rb $D_{2}$ transition ($5S_{1/2} \rightarrow 5P_{3/2}$, $\sim$780 nm,  $\Gamma_{0}/2\pi \approx 6$ MHz), $\Delta \sim 0.1$ corresponds to a frequency shift of $\sim 0.6$ MHz. Sub-MHz frequency stabilization has been experimentally demonstrated using Doppler-free saturated absorption spectroscopy \cite{utreja2022frequency}. For the strong $1S_{0} \rightarrow 1P_{1}$ transition of $^{174}$Yb ($\sim$ 399 nm, $\Gamma_{0}/2\pi \approx 29$ MHz), $\Delta \sim 0.1$ corresponds to a frequency shift of $\sim$ 2.9 MHz. Modulation transfer microscopy with balanced detection has achieved frequency instabilities at the kHz level, corresponding to a precision of $\sim 10^{-4}$ in $\Delta$ \cite{de2024laser}. Thus, the level of detuning required to achieve the directional switching demonstrated here is  well within experimental reach for typical strong electric-dipole transitions. 

\section{\MakeUppercase{ATOMIC TRIMER AS A MAGNETIC OPTICAL ANTENNA}}
\label{sec:emission}

At optical frequencies, the magnetic response of natural materials is weak because light-matter interactions are predominantly governed by electric dipole transitions. In recent decades, several systems, including nanoantennas, have been explored to suppress electric fields and enhance magnetic fields to strengthen magnetic transitions and increase the magnetic Purcell factor \cite{taminiau2012quantifying,karaveli2011spectral}. Atomic antennas supporting magnetic modes exhibit strong magnetic field and large decay-rate enhancement, providing a suitable platform for studying magnetic transitions in atoms \cite{alaee2020quantum}. Among the magnetic modes observed, mode 9 under in-plane $p$-polarized excitation induces a magnetic moment that is particularly interesting. The magnetic dipole is consistently oriented along the $z$-direction (Table \ref{tab:trimer_scat_s_pol}), nearly coincides with the incident excitation direction for small angles of incidence. This configuration causes the magnetic dipole radiation from the central atom to be emitted predominantly perpendicular to the incident excitation, allowing efficient signal collection while avoiding background scattering. Since the electric field is suppressed while  the magnetic field is strongly enhanced at the central atom and nearby regions, the detected signal is expected to carry signatures of magnetic transitions in the central atom. To quantify these effects, we evaluate the magnetic Purcell factor and the electric and magnetic field enhancements. 
	
The antiparallel dipole configuration of mode 9 in the linear trimer under $p$-polarized excitation leads to magnetic field enhancement in a region near the antenna center and electric field suppression there due to destructive interference. The antenna center therefore behaves as a magnetic hotspot, indicating a favorable environment for enhancing magnetic dipole (MD) transitions. To quantify this, we evaluate the magnetic Purcell factor $F_\text{M}$, the magnetic-to-electric field intensity contrast ratio (MF/EF), and the degree of magnetic emission (DME). The results are summarized in Table \ref{tab:angle_metrics}. The magnetic Purcell factor $F_\text{M}$ is proportional to the magnetic local density of states (LDOS) $\rho_\text{M}$ ($F_\text{M} \propto \rho_\text{M}$) \cite{horvath2023strong,novotny2012principles}. While $\rho_\text{M}$ is determined by the magnetic Green's function of the system, it is closely related to the local magnetic field intensity near resonance. Accordingly, $F_\text{M} = \Gamma_\text{M} / \Gamma_{0} \approx \rho_\text{M} / \rho_\text{M}^{0}$ can be approximated as $|\mathbf{H}_\text{tot}|^2 / |\mathbf{H}_\text{inc}|^2$, where $\Gamma_\text{M}$ is the MD decay-rate of the central atom, $\Gamma_{0}$ is the free-space MD decay-rate, and $\rho_\text{M}^{0}$ is the magnetic LDOS in free space~\cite{choi2016selective}. The calculated values for varying angles of incidence $\phi$ are shown in the second column of Table \ref{tab:angle_metrics}. The results show a significant enhancement in $F_\text{M}$, reaching $10^{3} - 10^{4}$ for all $\phi$. Further, to ensure that the region is magnetically dominated, we evaluate the magnetic-to-electric field intensity contrast ratio (MF/EF) using $\eta^{2} |\mathbf{H}_{\text{tot}}|^{2} / |\mathbf{E}_{\text{tot}}|^{2}$, where $\eta$ is the free-space impedance \cite{gangrskaia2025probing,martin2024optical}. The ratio exceeds unity for all angles of incidence $\phi \gtrsim 25^\circ$, as shown in the third column of Table \ref{tab:angle_metrics}, indicating a magnetically dominated region. 
Another  parameter that characterizes the extent of magnetic emission is the degree of magnetic emission (DME), defined as  $(\rho_\text{M}-\rho_{E})/(\rho_\text{M}+\rho_{E})$. It provides a measure of the relative modification of the electric and magnetic LDOS ($\rho_{E}$ and $\rho_\text{M}$, respectively) and lies in the range $-1 \leq DME \leq 1$. $\text{DME} < 0$ indicates enhancement of $\rho_{E}$, $\text{DME} \approx 0$ gives $\rho_{E} \approx \rho_\text{M}$, and $\text{DME} > 0$ indicates enhancement of $\rho_\text{M}$~\cite{sanz2018enhancing}. The evaluated DME for different angles of incidence $\phi$ are given in the fourth column of Table \ref{tab:angle_metrics}. Beyond $\phi \approx 25^\circ$, DME remains positive, indicating a sustained enhancement of magnetic LDOS and the formation of pronounced magnetic hotspots within the system. 

\begin{table}[h!]
\caption{Calculated values of the magnetic Purcell factor $F_\text{M}$, magnetic-to-electric field intensity contrast ratio (MF/EF), the degree of magnetic emission (DME), and the enhanced decay-rate of $\mathrm{Eu}^{3+}$ ($\Gamma_\text{M}$) for the linear trimer under in-plane $p$-polarized incidence, shown as a function of $\phi$ for mode 9 excitation.}
\label{tab:angle_metrics}
\begin{ruledtabular}
\begin{tabular}{ccccc}
$\phi$ & $F_\text{M}$ & MF/EF & DME & $\Gamma_\text{M}$ ($\mathrm{Eu}^{3+}$) \\
\colrule
$15^\circ$ & $2.2 \times 10^{3}$ & 0.36 & -0.48 & $3.2 \times 10^{4}$ \\
$30^\circ$ & $6.5 \times 10^{3}$ & 1.89 & 0.31 & $9.3 \times 10^{4}$ \\
$45^\circ$ & $1.2 \times 10^{4}$ & 7.07 & 0.75 & $1.7 \times 10^{5}$ \\
$60^\circ$ & $1.7 \times 10^{4}$ & 23.10 & 0.92 & $2.4 \times 10^{5}$\\
$75^\circ$ & $2.0 \times 10^{4}$ & 61.28 & 0.97 & $2.9 \times 10^{5}$\\
$90^\circ$ & $2.2 \times 10^{4}$ & 93.02 & 0.98 & $3.2 \times 10^{5}$\\
\end{tabular}
\end{ruledtabular}
\end{table}

As an example of a magnetic dipole emitter, $\mathrm{Eu}^{3+}$ exhibits a $^{5}D_{0} \rightarrow ^{7}F_{1}$ magnetic dipole transition at $584~\mathrm{nm}$, with a free-space radiative decay rate $\Gamma_{0} = 14.37~\mathrm{s}^{-1}$ \cite{dodson2012magnetic}. 
From the magnetic hotspot of the linear trimer, we obtain a decay-rate enhancement of $10^{3}$–$10^{4}$, implying that the MD transition rate of $\mathrm{Eu}^{3+}$ in this region, $\Gamma_\text{M} = F_\text{M} \Gamma_{0}$ reaches values on the order of $10^{4}$–$10^{5}~\mathrm{s}^{-1}$. 
This corresponds to a reduction of the excited-state lifetime from milliseconds to microseconds range. 
The calculated $\Gamma_\text{M}$ for $\mathrm{Eu}^{3+}$ at different angles of incidence are summarized in the last column of Table \ref{tab:angle_metrics}. 
Thus, the proposed system supports magnetic hotspots and operates as a magnetic optical antenna, evidenced by the large Purcell factor, magnetic field dominance, and positive DME over a broad angular range.

\section{\MakeUppercase{CONCLUSION}}  
\label{sec:conclusion}

We have presented a symmetry-resolved modal analysis of atomic trimers continuously tuned from linear ($D_{\infty h}$) to equilateral ($D_{3h}$), providing a complete map of collective modes and their evolution under controlled symmetry reduction. This framework reveals how symmetry governs mode excitation, degeneracy lifting, and access to the full modal spectrum, including a crossing between modes of the same irrep. Building on modal understanding, we demonstrate forward-backward scattering switching solely via frequency detuning, without requiring geometric reconfiguration. Furthermore, we show that a linear trimer under in-plane $p$-polarized excitation supports a magnetic mode with a strongly enhanced magnetic field and a large Purcell factor exceeding $10^4$, enabling efficient probing of magnetic dipole transitions through emission predominantly directed into the transverse plane. These results establish atomic trimers as a minimal platform where symmetry-controlled mode engineering enables dynamic scattering control and pronounced magnetic enhancement, with potential applications in tunable atomic-scale photonic systems and probing magnetic dipole transitions in atoms such as Eu$^{3+}$. The presented framework and switching mechanisms provide a foundation for designing reconfigurable atomic-scale light-manipulation devices, including directional emitters, sensors, and quantum optical interfaces. Specifically, the frequency‑controlled forward‑backward switching (Fig.~\ref{fig:Backscat_5}) provides an atomic‑scale optical router while the magnetic hotspot with Purcell factor $>10^4$ can selectively enhance magnetic dipole emission from quantum emitters such as NV centers or Eu$^{3+}$.

\section*{Acknowledgments}
The author M.A.P. acknowledges the financial support provided by CUSAT under the SMNRI Grant No. CUSAT/PL(UGC).A5/3860/2022 dated 25/08/2022. L.G.J. acknowledges a PhD grant from CUSAT. 

\section*{Data Availability}
The data that support the findings of this article are not publicly available. The data are available from the authors upon reasonable request.

\appendix
\renewcommand{\thesection}{\Alph{section}}

\section{Coupled-dipole framework and Green's tensor for the atomic trimer}
\label{App.A}
In this Appendix, we present the coupled-dipole expressions and the Green's tensor describing the self-consistent interactions among individual dipolar emitters in an atomic trimer antenna under $s$- and $p$-polarized plane wave excitation, as introduced in Sec. \ref{sec:theory}. The schematic of the system is shown in Fig. \ref{fig:Trimer_schematic}. The atomic trimer lies in the $xz$ plane. Accordingly, $s$-polarized excitation involves field components along $x$ and $z$, resulting in six coupled equations for the dipole components $\{p_{i}^{x}, p_{i}^{z}\}$ of the $i^{th}$ atom ($i = 1, 2, 3$), as given in Eqs. (\ref{eq:p1x}) - (\ref{eq:p3z}). In contrast, for $p$-polarized excitation, the incident field can have components along $x$, $y$ and $z$, depending on the incidence, yielding nine coupled equations $\{ p_{i}^{x}, p_{i}^{y},p_{i}^{z} \}$, as given in Eqs. (\ref{eq:p1x}) - (\ref{eq:p3y}). Solving these coupled equations yields the self-consistent induced dipole moments of each atom, under the respective incident polarizations. 
\begin{equation}\label{eq:p1x}
    p^{x}_{1} = \epsilon_{0} \alpha \left [E^{x}_\text{inc}(\mathbf{r}_{1}) + G^{xx}_{12} p^{x}_{2} + G^{xx}_{13} p^{x}_{3} + G^{xz}_{13} p^{z}_{3} \right ]
\end{equation}	
\begin{equation}\label{eq:p1z}
    p^{z}_{1} = \epsilon_{0} \alpha \left [ E^{z}_\text{inc}(\mathbf{r}_{1}) + G^{zz}_{12} p^{z}_{2} + G^{xz}_{13} p^{x}_{3} + G^{zz}_{13} p^{z}_{3} \right ] 
\end{equation}
\begin{equation}\label{eq:p2x}
    p^{x}_{2} = \epsilon_{0} \alpha \left [ E^{x}_\text{inc}(\mathbf{r}_{2}) + G^{xx}_{23} p^{x}_{3} + G^{xz}_{23} p^{z}_{3} + G^{xx}_{12} p^{x}_{1} \right ]
\end{equation}	
\begin{equation}\label{eq:p2z}
    p^{z}_{2} = \epsilon_{0} \alpha \left [ E^{z}_\text{inc}(\mathbf{r}_{2}) + G^{xz}_{23} p^{x}_{3} + G^{zz}_{23} p^{z}_{3} + G^{zz}_{12} p^{z}_{1}  \right ] 
\end{equation}
\begin{equation}\label{eq:p3x}
    p^{x}_{3} = \epsilon_{0} \alpha \left [ E^{x}_\text{inc}(\mathbf{r}_{3}) + G^{xx}_{13} p^{x}_{1} + G^{xz}_{13} p^{z}_{1} + G^{xx}_{23} p^{x}_{2} + G^{xz}_{23} p^{z}_{2}  \right ] 
\end{equation}	
\begin{equation}\label{eq:p3z}
    p^{z}_{3} = \epsilon_{0} \alpha \left [ E^{z}_\text{inc}(\mathbf{r}_{3}) + G^{xz}_{13} p^{x}_{1} + G^{zz}_{13} p^{z}_{1} + G^{xz}_{23} p^{x}_{2} + G^{zz}_{23} p^{z}_{2}  \right ]. 
\end{equation}
    \begin{equation}\label{eq:p1y}
    p^{y}_{1} = \epsilon_{0} \alpha \left [ E^{y}_\text{inc}(\mathbf{r}_{1}) + G^{yy}_{12} p^{y}_{2} + G^{yy}_{13} p^{y}_{3} \right ] 
\end{equation}
\begin{equation}\label{eq:p2y}
    p^{y}_{2} = \epsilon_{0} \alpha \left [ E^{y}_\text{inc}(\mathbf{r}_{2}) + G^{yy}_{23} p^{y}_{3} + G^{yy}_{12} p^{y}_{1} \right ]
\end{equation}
\begin{equation}\label{eq:p3y}
    p^{y}_{3} = \epsilon_{0} \alpha \left [ E^{y}_\text{inc}(\mathbf{r}_{3}) + G^{yy}_{13} p^{y}_{1} + G^{yy}_{23} p^{y}_{2} \right ] 
\end{equation}
Here, $\alpha = [-(\Gamma_{0}/2)\alpha_{0}]/[\delta + i (\Gamma_{0})/2]$ is the single-atom polarizability, where $\alpha_{0} = 6\pi/k^{3}$, $\Gamma_{0}$ is the radiative decay-rate, and $\delta = \omega_\text{inc} - \omega_{a}$ is the frequency detuning between the incident frequency $\omega_\text{inc}$ and the atomic transition frequency $\omega_{a}$. The quantities $E^{x}_\text{inc}(\mathbf{r}_{i})$, $E^{y}_{\text{inc}}(\mathbf{r}_{i})$ and  $E^{z}_{\text{inc}}(\mathbf{r}_{i})$ ($i=1,2,3$) are the components of the incident electric field at the position $\mathbf{r}_{i}$. The incident electric field is taken as a plane wave of the form $\mathbf{E}_{\text{inc}}(\mathbf{r})=E_{0} e^{-i \mathbf{k} \cdot \mathbf{r}} \mathbf{e}$, where $E_{0}$ is the field amplitude, $\mathbf{e}$ the polarization unit vector, and $\mathbf{k}$ is the wave vector given by $\mathbf{k} = k\left(\sin\phi \cos \psi\,\hat{\mathbf{x}} + \sin \psi\,\hat{\mathbf{y}}+ \cos\phi \cos\psi \,\hat{\mathbf{z}}\right)$, with $\phi$ the in-plane angle of incidence measured with respect to $z$-axis and $\psi$ the out-of-plane angle of incidence measured from trimer plane. $G$ denotes the Green's tensor of the trimer, whose components are detailed below. 

In general, the Green's tensor of a system is given by \cite{jackson1998classical,tai1971dyadic}
\begin{equation}
    \overline{\overline{\mathbf{G}}}(\mathbf{r}_{i},\mathbf{r}_{j})=\frac{3}{2\epsilon_{0}\alpha_{0}}e^{i\zeta}\left[g_{1}(\zeta)\,\overline{\overline{\mathbf{I}}}+g_{2}(\zeta)\,\mathbf{n}\mathbf{n}\right],
\end{equation}
where $\overline{\overline{\mathbf{I}}}$ is the identity dyadic, $\mathbf{n}=\frac{\mathbf{r}_{i}-\mathbf{r}_{j}}{\left|\mathbf{r}_{i}-\mathbf{r}_{j}\right|}$, and $\zeta=\left|k(\mathbf{r}_{i}-\mathbf{r}_{j})\right|$.
Using this, the Cartesian components of the generalized Green's tensor for an isosceles atomic trimer lying in the $xz$ plane can be derived as 
\begin{equation}
    \overline{\overline{\mathbf{G}}}_{12} =
    \begin{bmatrix}
        G_{12}^{xx} & G_{12}^{xy} & G_{12}^{xz}\\
        G_{12}^{yx} & G_{12}^{yy} & G_{12}^{yz}\\
        G_{12}^{zx} & G_{12}^{zy} & G_{12}^{zz}
    \end{bmatrix}
    =\frac{3}{2 \epsilon_{0} \alpha_{0}} e^{i \zeta} \begin{bmatrix}
        g_{1} + g_{2} & 0 & 0\\
        0 & g_{1} & 0\\
        0 & 0 & g_{1}
    \end{bmatrix},
\end{equation}  
\begin{equation}
    \overline{\overline{\mathbf{G}}}_{13} =
    \begin{bmatrix}
        G_{13}^{xx} & G_{13}^{xy} & G_{13}^{xz}\\
        G_{13}^{yx} & G_{13}^{yy} & G_{13}^{yz}\\
        G_{13}^{zx} & G_{13}^{zy} & G_{13}^{zz}
    \end{bmatrix}
    =\frac{3}{2 \epsilon_{0} \alpha_{0}} e^{i \zeta'} \begin{bmatrix}
        g'_{1} + (\frac{l'^{2}}{4 n}) g'_{2} & 0 & -\frac{m l'}{2 n} g'_{2}\\
        0 & g'_{1} & 0\\
        -\frac{m l'}{2 n} g'_{2} & 0 & g'_{1} + \frac{m^{2}}{n}g'_{2}
    \end{bmatrix},
\end{equation}  
\begin{equation}
    \overline{\overline{\mathbf{G}}}_{23} =
    \begin{bmatrix}
        G_{23}^{xx} & G_{23}^{xy} & G_{23}^{xz}\\
        G_{23}^{yx} & G_{23}^{yy} & G_{23}^{yz}\\
        G_{23}^{zx} & G_{23}^{zy} & G_{23}^{zz}
    \end{bmatrix}
    =\frac{3}{2 \epsilon_{0} \alpha_{0}} e^{i \zeta'} \begin{bmatrix}
        g'_{1} + (\frac{l'^{2}}{4 n}) g'_{2} & 0 & \frac{m l'}{2 n} g'_{2}\\
        0 & g'_{1} & 0\\
        \frac{m l'}{2 n} g'_{2} & 0 & g'_{1} + \frac{m^{2}}{n}g'_{2}
    \end{bmatrix}.
\end{equation}
Here,
\begin{equation}
\begin{aligned}
    g_{1}&=\frac{1}{\zeta}+\frac{i}{\zeta^{2}}-\frac{1}{\zeta^{3}}, \qquad g_{2}=-\frac{1}{\zeta}-\frac{3i}{\zeta^{2}}+\frac{3}{\zeta^{3}},\\
    g'_{1}&=\frac{1}{\zeta'}+\frac{i}{\zeta'^{2}}-\frac{1}{\zeta'^{3}}, \qquad g'_{2}=-\frac{1}{\zeta'}-\frac{3i}{\zeta'^{2}}+\frac{3}{\zeta'^{3}},
\end{aligned}
\end{equation}
with $\zeta = kl'$, $\zeta' = k \sqrt{n}$, $n = \frac{l'^{2}}{4} + m^{2}$, and $m = l \sin \theta$. The parameters $l$, $l'$, $m$, and $\theta$ are marked in Fig. \ref{fig:Trimer_schematic}(a). 

\section{Eigenmode decomposition of the atomic trimer}
\label{App.B}
In this Appendix, we discuss the expressions required to study the response of an atomic trimer antenna using eigenmode decomposition approach, as introduced in Sec. \ref{sec:EMD}. 
The interactions within the antenna are governed by the matrix equation $\mathbf{E}_{\text{inc}}=\mathbf{M} \mathbf{p}$ \cite{foldy1945multiple,mulholland1994light}, where $\mathbf{E}_{\text{inc}}$ is the incident field vector, $\mathbf{M}$ is the coupling matrix, and $\mathbf{p}$ is the vector of individual dipole moments. 
For a generalized atomic trimer, this yields a $9 \times 9$ system (see Eq. (\ref{eq:EMD_matrix})), whose decomposition gives nine eigenvalues and nine corresponding eigenvectors, indicating that an atomic trimer supports a total of nine eigenmodes.
The matrix equation for the trimer lying in $xz$ plane, incorporating both $s$- and $p$-polarized excitations, can be written as

\begin{equation}\label{eq:EMD_matrix}
    \begin{bmatrix}
        E_\text{inc}^{x}(\mathbf{r}_{1}) \\ E_\text{inc}^{y}(\mathbf{r}_{1}) \\ E_{\text{inc}}^{z}(\mathbf{r}_{1}) \\ E_\text{inc}^{x}(\mathbf{r}_{2}) \\ E_\text{inc}^{y}(\mathbf{r}_{2}) \\ E_\text{inc}^{z}(\mathbf{r}_{2}) \\ E_{\text{inc}}^{x}(\mathbf{r}_{3}) \\ E_{\text{inc}}^{y}(\mathbf{r}_{3}) \\ E_{\text{inc}}^{z}(\mathbf{r}_{3})
    \end{bmatrix}
    =
    \begin{bmatrix}
        \alpha^{-1} & 0 & 0 & -G_{12}^{xx} & 0 & 0 & -G_{13}^{xx} & 0 & -G_{13}^{xz} \\ 
        0 & \alpha^{-1} & 0 & 0 & -G_{12}^{yy} & 0 & 0 & -G_{13}^{yy} & 0 \\ 
        0 & 0 & \alpha^{-1} & 0 & 0 & -G_{12}^{zz} & -G_{13}^{zx} & 0 & -G_{13}^{zz} \\
        -G_{12}^{xx} & 0 & 0 & \alpha^{-1} & 0 & 0 & -G_{23}^{xx} & 0 & -G_{23}^{xz} \\
        0 & -G_{12}^{yy} & 0 & 0 & \alpha^{-1} & 0 & 0 & -G_{23}^{yy} & 0 \\
        0 & 0 & -G_{12}^{zz} & 0 & 0 & \alpha^{-1} & -G_{23}^{zx} & 0 & -G_{23}^{zz} \\
        -G_{13}^{xx} & 0 & -G_{13}^{xz} & -G_{23}^{xx} & 0 & -G_{23}^{xz} & \alpha^{-1} & 0 & 0 \\
        0 & -G_{13}^{yy} & 0 & 0 & -G_{23}^{yy} & 0 & 0 & \alpha^{-1} & 0 \\
        -G_{13}^{zx} & 0 & -G_{13}^{zz} & -G_{23}^{zx} & 0 & -G_{23}^{zz} & 0 & 0 & \alpha^{-1}
    \end{bmatrix}
    \begin{bmatrix}
        p_{1}^{x} \\ p_{1}^{y} \\ p_{1}^{z} \\ p_{2}^{x} \\ p_{2}^{y} \\ p_{2}^{z} \\ p_{3}^{x} \\ p_{3}^{y} \\ p_{3}^{z}
    \end{bmatrix}.
\end{equation}

Solving the above matrix equation yields the induced dipole moment components of each atom, from which the magnitude and orientation of the excited modes can be determined.

The total scattering cross section ($\sigma_{sc}$) of the antenna, as obtained from the eigenmode decomposition is given by \cite{bettles2016cooperative}
\begin{equation} \label{eq:EMD_Appendix}
    \sigma_{sc} = \frac{\sigma_{0}}{\alpha_{0} |E_\text{inc}|^{2}} \left [\sum_{n} |b_{n}|^{2} \, \text{Im} \left (\frac{1}{\mu_{n}}\right ) + \sum_{n,n'}^{n \neq n'} \text{Im}\left(\frac{b_{n}^{*} b_{n'}}{\mu_{n'}} \, \mathbf{m}_{n}^{*} \cdot \mathbf{m}_{n'}\right) \right].
\end{equation}
Here, $\sigma_{0} = 6 \pi/k^{2}$ is the resonant atomic scattering cross section, and $\mu_{n}$ and $\mathbf{m}_{n}$ are the eigenvalues and eigenvectors, respectively, of the coupling matrix $\mathbf{M}$, with $n$ the eigenmode index. The coefficient $b_{n}$ is obtained by further solving a set of coupled linear equations given by 
\begin{equation} \label{eq:bCoeff_equation}
    \mathbf{m}_{n}^{*} \cdot \mathbf{E}_\text{inc} = b_{n} + \sum_{n' \neq n} b_{n'} \, \mathbf{m}_{n}^{*} \cdot \mathbf{m}_{n'}.
\end{equation}

\bibliographystyle{apsrev4-2}
\bibliography{references.bib}

\end{document}